\documentclass[letterpaper,11pt]{article}
\pdfoutput=1
\pdfoutput=1

\usepackage{jheppub}
\usepackage{multirow}

\usepackage{subfig}
\usepackage{xspace}
\usepackage[countmax]{subfloat}

\usepackage[normalem]{ulem}
\usepackage{amssymb}
\usepackage{amsmath}
\usepackage{cancel}
\usepackage{color}
\usepackage{braket}
\usepackage{graphicx}
\usepackage{multirow}
\usepackage{verbatim}
\usepackage{amsthm}
\usepackage{slashed}
\usepackage{wasysym}
\usepackage{simplewick}
\usepackage{mathtools}
\usepackage{soul}
\usepackage{xspace}

\newcommand{\Tau}{\mathcal{T}}

\newcount\colveccount
\newcommand*\colvec[1]{
        \global\colveccount#1
        \begin{pmatrix}
        \colvecnext
}
\def\colvecnext#1{
        #1
        \global\advance\colveccount-1
        \ifnum\colveccount>0
                \\
                \expandafter\colvecnext
        \else
                \end{pmatrix}
        \fi
}

\newcommand{\fd}[2]{\parbox{#1}{\includegraphics[width=#1]{#2}}}

\newcommand{\EEC}{{\text{EEC}}}

\newcommand{\z}{\zeta}
\newcommand{\bz}{\bar z}

\newcommand{\nt}{\notag\\}

\def\cG{\mathcal{G}}

\def\cL{\mathcal{L}}
\def\cM{\mathcal{M}}
\def\cN{\mathcal{N}}
\def\cO{\mathcal{O}}
\def\cS{\mathcal{S}}

\def\cP{\mathcal{P}}

\def\cY{\mathcal{Y}}

\def\tr{{\rm tr}}

\newcommand{\dbar}{d\hspace*{-0.08em}\bar{}\hspace*{0.1em}}

\newcommand{\kperp}{{\vec{k}_\perp^{\,2}}}

\newcommand{\dfslash}{{\df\!\!\!{}^-}}
\newcommand{\kperpeps}{{\vec{k}_\perp^{\,2\epsilon}}}

\def\nn{{\nonumber}}

\newcommand{\erfi}{\mathrm{erfi}}
\newcommand{\erf}{\mathrm{erf}}

\newcommand{\sceti}{SCET$_{\rm I}$}
\newcommand{\scetii}{SCET$_{\rm II}$}

\newcommand{\Eq}[1]{Equation~\eqref{#1}}

\DeclareRobustCommand{\Sec}[1]{Sec.~\ref{#1}}

\DeclareRobustCommand{\Fig}[1]{Fig.~\ref{#1}}

\DeclareRobustCommand{\Eq}[1]{Eq.~(\ref{#1})}

\def\be{\begin{equation}}
\def\ee{\end{equation}}

\allowdisplaybreaks[3]

\newcommand{\eq}[1]{Eq.~\eqref{eq:#1}}

\newcommand{\ord}[1]{\mathcal{O}(#1)}

\newcommand{\df}{\mathrm{d}}
\newcommand{\img}{\mathrm{i}}

\newcommand{\gcusp}{\Gamma^{\mathrm{cusp}}}

\newcommand\bn{{\bar n}}

\usepackage{marginnote}

  \newcount\hour \newcount\minute
  \hour=\time \divide \hour by 60 \minute=\time
  \newcommand{\todaytime}{\today \ -- \number\hour :\ifnum \minute<10 0\fi\number\minute}

\preprint{MIT-CTP 5162, MPP-2019-243}

\title{Subleading Power Resummation of Rapidity Logarithms:\\The Energy-Energy Correlator in $\cN=4$ SYM}

\abstract{We derive and solve renormalization group equations that allow for the resummation of subleading power rapidity logarithms. Our equations involve operator mixing into a new class of operators, which we term the ``rapidity identity operators", that will generically appear at subleading power in problems involving both rapidity and virtuality scales. To illustrate our formalism, we analytically solve these equations to resum the power suppressed logarithms appearing in the back-to-back (double light cone) limit of the Energy-Energy Correlator (EEC) in $\cN$=4 super-Yang-Mills. These logarithms can also be extracted to $\cO(\alpha_s^3)$ from a recent perturbative calculation, and we find perfect agreement to this order. Instead of the standard Sudakov exponential,  our resummed result for the subleading power logarithms is expressed in terms of Dawson's integral, with an argument related to the cusp anomalous dimension. We call this functional form ``Dawson's Sudakov". Our formalism is widely applicable for the resummation of subleading power rapidity logarithms in other more phenomenologically relevant observables, such as the EEC in QCD, the $p_T$ spectrum for color singlet boson production at hadron colliders, and the resummation of power suppressed logarithms in the Regge limit. 
 }

\begin{document} 

\author[\eta]{Ian Moult,}
\emailAdd{imoult@slac.stanford.edu}
\author[\mu]{Gherardo Vita}
\emailAdd{vita@mit.edu}
\author[\nu]{and Kai Yan}
\emailAdd{kyan@mpp.mpg.de}

\affiliation[\eta]{SLAC National Accelerator Laboratory, Stanford University, CA, 94309, USA\vspace{0.5ex}}
\affiliation[\mu]{Center for Theoretical Physics, Massachusetts Institute of Technology, Cambridge, MA 02139, USA}
\affiliation[\nu]{Max-Planck-Institut fur Physik, Werner-Heisenberg-Institut, 80805 Munchen, Germany\vspace{0.5ex}}

\maketitle
\section{Introduction}\label{sec:intro}

Scattering amplitudes and cross sections simplify in infrared kinematic limits enabling insight into their all orders perturbative structure.
One of the most powerful approaches to studying the all orders structure is the use of renormalization group (RG) techniques. Depending on the particular nature of the kinematic limit considered, these RG equations often describe evolution not just in the standard virtuality scale, $\mu$, but also in other additional physical scales. A common example in gauge theories are rapidity evolution equations, which allow for the resummation of infrared logarithms associated with hierarchical scales in rapidity. Classic examples include the Sudakov form factor \cite{Collins:1989bt},  the Collins-Soper equation \cite{Collins:1981uk,Collins:1981va,Collins:1984kg} describing the $p_T$ spectrum for color singlet boson production in hadron colliders, the BFKL evolution equations describing the Regge limit  \cite{Kuraev:1977fs,Balitsky:1978ic,Lipatov:1985uk},  and the rapidity renormalization group \cite{Chiu:2011qc,Chiu:2012ir} describing more general event shape observables.

There has recently been significant effort towards understanding subleading power corrections to infrared limits  \cite{Manohar:2002fd,Beneke:2002ph,Pirjol:2002km,Beneke:2002ni,Bauer:2003mga,Hill:2004if,Lee:2004ja,Dokshitzer:2005bf,Trott:2005vw,Laenen:2008ux,Laenen:2008gt,Paz:2009ut,Benzke:2010js,Laenen:2010uz,Freedman:2013vya,Freedman:2014uta,Bonocore:2014wua,Larkoski:2014bxa,Bonocore:2015esa,Bonocore:2016awd,Kolodrubetz:2016uim,Moult:2016fqy,Boughezal:2016zws,DelDuca:2017twk,Balitsky:2017flc,Moult:2017jsg,Goerke:2017lei,Balitsky:2017gis,Beneke:2017ztn,Feige:2017zci,Moult:2017rpl,Chang:2017atu,Beneke:2018gvs,Beneke:2018rbh,Moult:2018jjd,Ebert:2018lzn,Ebert:2018gsn,Bhattacharya:2018vph,Boughezal:2018mvf,vanBeekveld:2019prq,vanBeekveld:2019cks,Bahjat-Abbas:2019fqa,Beneke:2019kgv,Boughezal:2019ggi,Moult:2019mog,Beneke:2019mua,Cieri:2019tfv,Moult:2019uhz,Beneke:2019oqx} with the ultimate goal of achieving a systematic expansion, much like for problems where their exists a local operator product expansion (OPE). Using Soft Collinear Effective Theory (SCET) \cite{Bauer:2000ew, Bauer:2000yr, Bauer:2001ct, Bauer:2001yt} subleading power infrared logarithms were resummed to all orders using RG techniques for a particular class of event shape observables where only virtuality evolution is required. This was achieved both in pure Yang-Mills theory \cite{Moult:2018jjd}, and including quarks in $\cN=1$ QCD \cite{Moult:2019uhz}, and a conjecture for the result including quarks in QCD was presented in \cite{Moult:2019uhz}. Subleading power infrared logarithms have also been resummed for color singlet production at kinematic threshold, when only soft real radiation is present \cite{Beneke:2018gvs,Beneke:2019mua,Bahjat-Abbas:2019fqa}.

In this paper we build on the recent advances in understanding the structure of subleading power renormalization group equations in SCET, and consider for the first time the resummation of subleading power rapidity logarithms. Using renormalization group consistency arguments, we derive a class of subleading power rapidity evolution equations. These equations involve mixing into new class of operators, which play a crucial role in the renormalization group equations. We call these operators ``rapidity identity operators", and we derive their renormalization group properties, and solve the associated RG equations. 

We apply our evolution equations to derive the all orders structure of the power suppressed leading logarithms for the energy-energy correlator (EEC) event shape in $\cN=4$ super-Yang-Mills (SYM) theory in the back-to-back (double light cone) limit. Denoting these power suppressed contributions by $\text{EEC}^{(2)}$, we find the remarkably simple formula
\begin{align}
\boxed{\text{EEC}^{(2)}=-\sqrt{2a_s}~D\left[ \sqrt{\frac{\gcusp}{2}} \log(1-z) \right]}\,,
\end{align}
where $D(x)=1/2\sqrt{\pi}e^{-x^2}\erfi(x)$ is Dawson's integral, $a_s=\alpha_s/(4\pi)C_A$, and $\gcusp$ is the cusp anomalous dimension \cite{Korchemsky:1987wg}. This result provides insight into new all orders structures appearing in subleading power infrared limits. Since this extends the classic Sudakov exponential \cite{Sudakov:1954sw}, we will refer to this functional form as ``Dawson's Sudakov". The particular example of the EEC observable was chosen in this paper, since its exact structure for generic angles is known to $\cO(\alpha_s^3)$ due to the remarkable calculation of \cite{Henn:2019gkr}, and we find perfect agreement with the expansion of their results in the back-to-back limit to this order, providing a strong check of our techniques. While we focus on the EEC in $\cN=4$, this observable has an identical resummation structure to $p_T$ resummation, and therefore  the techniques we have developed apply more generally, both to the EEC in QCD, and to the $p_T$ distribution of color singlet bosons at hadron colliders.

An outline of this paper is as follows. In \Sec{sec:EEC_b2b} we review the known structure of the EEC observable in the back-to-back limit, and relate it to the case of the $p_T$ spectrum of color singlet bosons, which is perhaps more familiar to the resummation community. In \Sec{sec:FO} we perform a fixed order calculation of the EEC at subleading power in SCET, which allows us to understand the structure of the subleading power rapidity divergences, and provides the boundary data for our RG approach. In \Sec{sec:RRG_NLP} we study the structure of subleading power rapidity evolution equations, introduce the rapidity identity operators, and analytically solve their associated evolution equations. In \Sec{sec:N4} we apply these evolution equations to the particular case of the EEC in $\cN=4$ SYM to derive the subleading power leading logarithmic series, and we comment on some of the interesting features of the result. We also compare our result with the fixed order calculation of \cite{Henn:2019gkr} expanded in the back-to-back limit, finding perfect agreement.  We conclude in \Sec{sec:conc}, and discuss many directions for improving our understanding of the infrared properties of gauge theories at subleading powers. 

\section{The Energy-Energy Correlator in the Back-to-Back Limit}\label{sec:EEC_b2b}

In this section we introduce the EEC observable, and review its structure in the back-to-back limit at leading power. We then discuss the resummation of the associated infrared logarithms using the rapidity renormalization group approach. This will allow us to introduce our notation, before proceeding to subleading power. 

An additional goal of this section is to make clear the relation between the EEC in the back-to-back limit and more standard $p_T$ resummation, which may be more familiar to the resummation community. This should also make clear that the techniques we develop are directly applicable to the case of $p_T$ resummation, although we leave a complete treatment of $p_T$ resummation to a future publication due to complexities related to the treatment of the initial state hadrons. Some other recent work towards understanding subleading power factorization for $p_T$ can be found in \cite{Balitsky:2017flc,Balitsky:2017gis}.

For a color singlet source, the EEC is defined as \cite{Basham:1978bw}
\begin{align}\label{eq:EEC_intro}
\text{EEC}(\chi)=\sum\limits_{a,b} \int d\sigma_{V\to a+b+X} \frac{2 E_a E_b}{Q^2 \sigma_{\text{tot}}}   \delta(\cos(\theta_{ab}) - \cos(\chi))\,,
\end{align}
where the sum is over all pairs of final state particles, $E_a$ are the energies of the particles, and $\theta_{ab}$ are the angles between pairs of particles. Energy correlators are a theoretically nice class of event shape observable, since they can be directly expressed in terms of energy flow operators \cite{Hofman:2008ar,Belitsky:2013xxa,Belitsky:2013bja,Belitsky:2013ofa}
\begin{align}
\mathcal{E}(\vec n) =\int\limits_0^\infty dt \lim_{r\to \infty} r^2 n^i T_{0i}(t,r \vec n)\,.
\end{align}
In particular, the EEC is given by the four-point Wightman correlator
\begin{align}
\frac{1}{\sigma_{\rm tot}} \frac{d\sigma}{dz}=\frac{\langle \cO \mathcal{E}(\vec n_1)  \mathcal{E}(\vec n_2) \cO^\dagger \rangle }{\langle \cO  \cO^\dagger \rangle}\,,
\end{align}
where we have introduced the convenient variable
\begin{align}
z=\frac{1-\cos {\chi}}{2}\,,
\end{align}
and $\cO$ is a source operator that creates the excitation.

There has been significant recent progress in understanding the EEC, both in QCD, as well as in conformal field theories. In QCD, the EEC has been computed for arbitrary angles at next-to-leading order (NLO) analytically \cite{Dixon:2018qgp,Luo:2019nig} and at NNLO numerically \cite{DelDuca:2016csb,DelDuca:2016ily}. In $\cN=4$ it has been computed for arbitrary angles to NNLO \cite{Belitsky:2013ofa,Henn:2019gkr}. There has also been significant progress in understanding the limits of the EEC, namely the collinear ($z\to 0$) limit \cite{Dixon:2019uzg,Kravchuk:2018htv,Kologlu:2019bco,Kologlu:2019mfz,Korchemsky:2019nzm}, and the back-to-back ($z\to 1$) limit \cite{deFlorian:2004mp,Moult:2018jzp,Korchemsky:2019nzm,Gao:2019ojf}. Here we will focus on the EEC in the back-to-back limit, where it exhibits Sudakov double logarithms. As we will explain shortly, these double logarithms are directly related to those appearing for transverse momentum resummation. In this section we follow closely the factorization derived in \cite{Moult:2018jzp} using the rapidity renormalization group \cite{Chiu:2012ir,Chiu:2011qc}. An alternative approach to studying this limit directly from the four point correlator was given in \cite{Korchemsky:2019nzm}.

\begin{figure}
\begin{center}
\includegraphics[width=0.75\columnwidth]{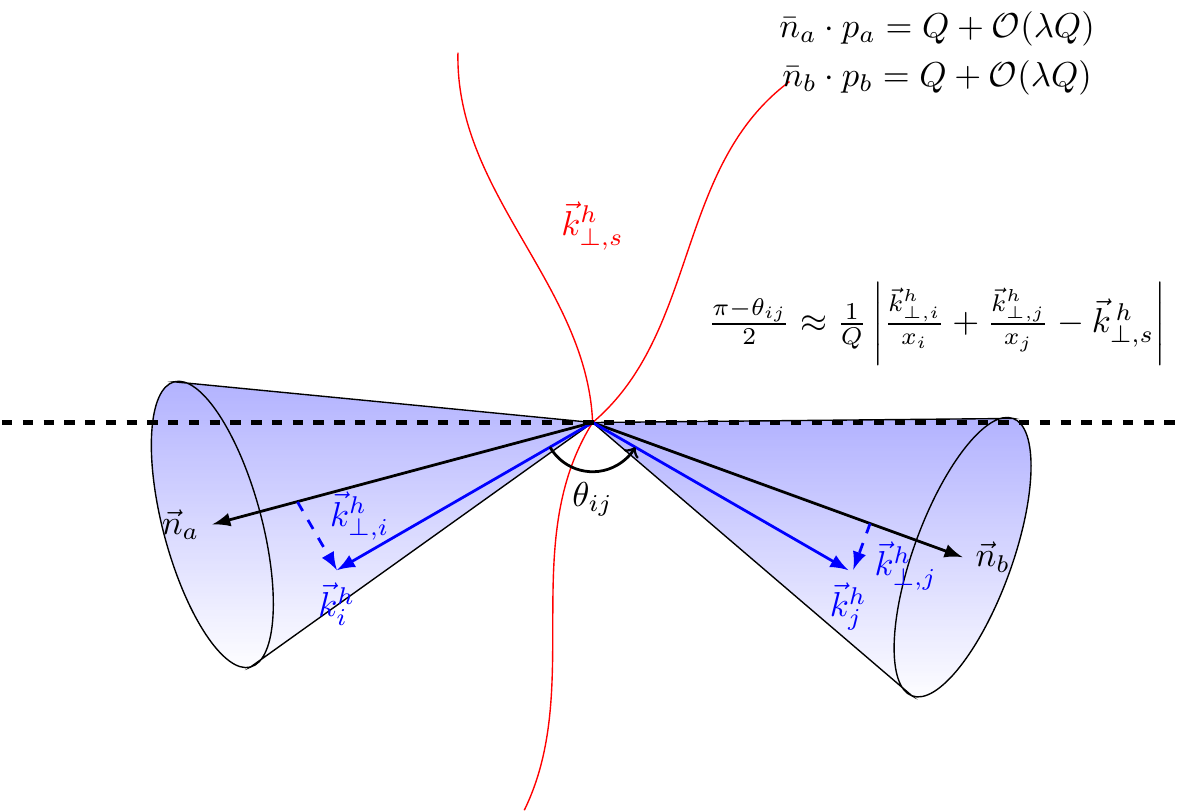}
\end{center}
\caption{The kinematics of the EEC in the back-to-back limit. Wide angle soft radiation recoils the two jets in a manner identical to the case of $p_T$ for color singlet boson production in hadronic collisions. This provides a precise relation between the factorization formulas in the two cases. (Figure from \cite{Moult:2018jzp}.)  }
\label{fig:schematic_factorization}
\end{figure}

The back-to-back limit corresponds to the region of phase space where there are two collimated jets accompanied by low energy soft radiation, which recoils the directions of the jets so that they are not exactly back-to-back. This configuration is illustrated in \Fig{fig:schematic_factorization}. A simple exercise shows that the angle between two partons correlated by the EEC is related to the transverse momentum of these particles within the jets and the transverse momentum of the soft radiation that recoils these jets, by
\begin{align}
\label{eq:z_in_back_to_back}
1-z=\frac{1}{Q^2}  \left|  \frac{\vec k_{\perp,i}^h}{x_i}+\frac{\vec k_{\perp,j}^h}{x_j}-\vec k_{\perp,s}^h \right|^2 +\cO(1-z)\,.
\end{align}
Here $x_i=2E_i/Q$, where $Q$ is the center of mass energy.
This relation makes clear the connection between the EEC in the back-to-back limit and transverse momentum resummation, which we will shortly extend to subleading powers.

To describe the EEC in this limit, we use an effective field theory description of the soft and collinear dynamics, where the relevant modes have the scalings
\begin{align}\label{eq:pc_modes}
p_s\sim Q(\lambda, \lambda, \lambda)\,, \qquad p_c \sim Q(\lambda^2, 1, \lambda) \,, \qquad p_{\bar c} \sim Q(1,\lambda^2,  \lambda)\,.
\end{align}
Here $\lambda$ is a power counting parameter and is defined as
\begin{align}
\lambda\sim \sqrt{1-z}\,.
\end{align}
This scaling defines what is referred to as an \scetii~ theory \cite{Bauer:2002aj}. Crucially, unlike in \sceti, the soft and collinear modes in \scetii~ have the same virtualities, but different rapidities. This factorization into soft and collinear modes therefore introduces divergences as $k^+/k^-\to \infty$ or $k^+/k^-\to 0$  \cite{Collins:1992tv,Manohar:2006nz,Collins:2008ht,Chiu:2012ir,Vladimirov:2017ksc}, which are referred to as rapidity divergences. To regulate these divergences, one must introduce a regulator that breaks boost invariance, allowing the soft and collinear modes to be distinguished. Once such a regulator is introduced, renormalization group evolution equations can be derived to resum the associated rapidity logarithms \cite{Chiu:2012ir,Chiu:2011qc}.

Using the effective field theory, one can systematically expand the cross section for either transverse momentum, or for the EEC in powers of the observable. For the EEC, we write
\begin{align}\label{eq:xsec_expand}
\frac{\df \sigma}{\df z}
&= \frac{\df\sigma^{(0)}}{\df z} + \frac{\df\sigma^{(2)}}{\df z}+ \frac{\df\sigma^{(4)}}{\df z} + \dotsb\nn \\
&=\text{EEC}^{(0)}+\text{EEC}^{(2)}+\text{EEC}^{(4)}\,,
\end{align}
where we will occasionally use the second notation, since it is more compact.
Here
\begin{align}
\frac{\df\sigma^{(0)}}{\df z}
&\sim \delta(1-z)+ \biggl[\frac{ \ord{1} }{1-z}\biggr]_+
\,, 
\end{align}
is referred to as the leading power cross section, and
describes all terms scaling like $\cO((1-z)^{-1})$ modulo logarithms. All the other terms in the expansion of the cross section are suppressed by explicit powers of $(1-z)$
\begin{align}\label{eq:scaling_lam2}
\frac{\df\sigma^{(2k)}}{\df z} &\sim \ord{(1-z)^{k-1}}
\,.\end{align}
The focus of this paper will be on deriving the structure of the leading logarithms in $\df\sigma^{(2)}/\df z$, which is also referred to as the next-to-leading power (NLP) cross section.

For the leading power cross section, $d\sigma^{(0)}/dz$, one can derive a factorization formula describing in  a factorized manner the contributions of the soft and collinear modes to the EEC in the $z\to 1$ limit \cite{Moult:2018jzp}
\begin{equation}\label{eq:fact_final}
\hspace{-0.35cm}\frac{d\sigma^{(0)}}{dz}= \frac{1}{2}  \int d^2 \vec k_\perp \int \frac{d^2 \vec b_\perp}{(2 \pi)^2} e^{-i \vec b_\perp \cdot \vec k_\perp} H(Q,\mu)  J^q_\EEC (\vec b_\perp,\mu,\nu) J^{\bar q}_\EEC (\vec b_\perp,\mu,\nu) S_\EEC(\vec b_\perp,\mu,\nu)  \delta \left( 1-z- \frac{\vec k_\perp^2}{Q^2} \right )\,,
\end{equation}
in terms of a hard function, $H$, jet functions, $J$, and a soft function, $S$. This factorization is nearly equivalent to the factorization for the $p_T$ for color singlet boson production (This factorization formula was originally derived in  \cite{Collins:1981uk,Collins:1981va,Collins:1984kg}, and was derived in terms of the rapidity renormalization group used here in \cite{Chiu:2012ir,Chiu:2011qc}), 
\begin{align}\label{eq:pt_fact}
\frac{1}{\sigma} \frac{d^3 \sigma^{(0)}}{d^2 \vec p_T dY dQ^2} = H(Q,\mu) \int \frac{d^2 \vec b_\perp}{(2\pi)^2} e^{i\vec b_\perp \cdot \vec p_T} \left[  B \times B  \right] (\vec b_\perp, \mu, \nu) S_\perp(\vec b_\perp, \mu, \nu)\,,
\end{align}
up to the fact that the jet functions are moved to the initial state, where they are referred to as beam functions \cite{Stewart:2009yx}. Apart from our intrinsic interest in understanding the kinematic limits of the EEC, this relation between the EEC and $p_T$ is one of our primary motivations for studying the EEC. Lessons derived from the EEC can be directly applied to understanding the structure of subleading power logarithms for $p_T$, which is a phenomenologically important observable at the LHC, for example, for precision studies of the Higgs boson. Here we briefly discuss the objects appearing in the factorization formula, both to emphasize the close connections between the EEC and $p_T$, as well as to introduce the general structure of the $\mu$ and $\nu$ rapidity evolution equations.

The hard functions, $H(Q,\mu)$, appearing in the factorization formulas for the EEC and $p_T$ are identical. They describe hard virtual corrections, and satisfy a multiplicative renormalization group equation (RGE) in $\mu$
\begin{align}
\mu \frac{d}{d\mu} H(Q,\mu) =2 \left[\gcusp(\alpha_s) \ln\frac{Q^2}{\mu^2} +  \gamma^H(\alpha_s) \right] H(Q,\mu)\,.
\end{align}
They are independent of rapidity.  The soft functions appearing in both $p_T$ and the EEC can be proven to be identical \cite{Moult:2018jzp}. They are matrix elements of Wilson lines, which for quarks and gluons are defined as
\begin{align}
 S_q(\vec p_T) &= \frac{1}{N_c} \big\langle 0 \big| \mathrm{Tr} \bigl\{
   \mathrm{T} \big[S^\dagger_{\bn} S_n\big]
   \delta^{(2)}(\vec p_T-\cP_\perp)\overline{\mathrm{T}} \big[S^\dagger_{n} S_{\bn} \big]
   \bigr\{ \big| 0 \big\rangle
\,,\nn\\
 S(\vec p_T) &= \frac{1}{N_c^2 - 1} \big\langle 0 \big| \mathrm{Tr} \bigl\{
   \mathrm{T} \big[\cS^\dagger_{\bn} \cS_n\big]
  \delta^{(2)}(\vec p_T-\cP_\perp) \overline{\mathrm{T}} \big[\cS^\dagger_{n} \cS_{\bn} \big]
   \bigr\} \big| 0 \big\rangle
\,.\end{align}
Here $\mathrm{T}$ and $\overline{\mathrm{T}}$ denote time and anti-time ordering, and $S_n$ and $\cS_n$ denote Wilson lines in the fundamental and adjoint representations, respectively. Explicitly, 
\begin{align}
 S_n(x) &= \mathrm{P} \exp\biggl[ \img g \int_{-\infty}^0 \df s\, n \cdot A(x+sn)\biggr]\,,
\end{align}
and similarly for the adjoint Wilson lines. These soft functions satisfy the $\mu$ and $\nu$ RGEs
\begin{align}
\nu \frac{d}{d\nu}S(\vec p_T)&=\int d\vec q_T \gamma^S_\nu(p_T-q_T) S(\vec q_T)\,, \nn \\
\mu \frac{d}{d\mu}S(\vec p_T)&=\gamma_\mu^S S(\vec p_T)\,,
\end{align}
with the anomalous dimensions
\begin{align}
\gamma_\mu^S &=4\gcusp(\alpha_s)\log \left( \frac{\mu}{\nu}\right)\,,\nn \\
\gamma_\nu^S &=2\gcusp (\alpha_s)\cL_0\left( \vec p_T,\mu \right)\,.
\end{align}
Here the color representation is implicit in the cusp anomalous dimension, and $\cL_0$ is a standard plus function (see e.g. \cite{Ebert:2016gcn} for a detailed discussion of two-dimensional plus distributions, and for a definition of the conventions that are used here). Since we will ultimately be interested in $\cN=4$ where all particles are in the same representation, we will drop the quark and gluon labels on the soft functions.

The only difference between $p_T$ and the EEC lies in the collinear sector, namely whether one uses beam functions or jet functions. The jet functions for the EEC were recently computed to NNLO  \cite{Luo:2019bmw,Luo:2019hmp}. Since in this paper we will be focused on resummation at LL, we can always choose to run all functions to the jet scale, and thereby avoid a discussion of the collinear sector for simplicity. The structure of the power corrections to the beam (jet) functions, and their matching to the parton distributions (fragmentation functions) is interesting, and will be presented in future work, since it is important for a complete understanding of $p_T$ at subleading powers.

These renormalization group evolution equations in both $\mu$ and $\nu$ allow for a derivation of the all orders structure of logarithms in the $z\to 1$ limit, at leading order in the $(1-z)$ expansion. Performing the renormalization group evolution, one finds for a non-conformal field theory \cite{Moult:2018jzp} (i.e. allowing for a running coupling)
\begin{align}
  \label{eq:resformula}
\frac{d\sigma^{(0)}}{dz} = &\; \frac{1}{4} \int\limits_0^\infty db\, b
  J_0(bQ\sqrt{1-z})H(Q,\mu_h) j^q_\EEC(b,b_0/b,Q) j^{\bar
  q}_\EEC(b,b_0/b,Q) S_\EEC( b,\mu_s, \nu_s) 
\nn\\
&\; \cdot
  \left(\frac{Q^2}{\nu_s^2}\right)^{\gamma^r_\EEC(\alpha_s(b_0/b))}
  \exp \left[ \int\limits_{\mu_s^2}^{\mu_h^2}
  \frac{d\bar{\mu}^2}{\bar{\mu}^2} \gcusp(\alpha_s(\bar \mu)) \ln
  \frac{b^2\bar{\mu}^2}{b_0^2} \right.
\nn\\
&\;
\left. +
  \int\limits_{\mu_h^2}^{b_0^2/b^2}\frac{d\bar{\mu}^2}{\bar{\mu}^2}
  \left(\gcusp(\alpha_s(\bar \mu)) \ln\frac{b^2 Q^2}{b_0^2} +
   \gamma^H (\alpha_s(\bar \mu)) \right) -
  \int\limits_{\mu_s^2}^{b_0^2/b^2}\frac{d\bar{\mu}^2}{\bar{\mu}^2}
   \gamma^s_\EEC (\alpha_s(\bar \mu))  \right]\,.
\end{align}
For the particular case of a conformal theory, this expression simplifies considerably, both due to the fact that the coupling doesn't run, and also since in a conformal field theory there is an equivalence between the rapidity anomalous dimension and the soft anomalous dimension  \cite{Vladimirov:2016dll,Vladimirov:2017ksc}. Combining this equivalence with the relations for the soft anomalous dimension derived in \cite{Dixon:2008gr} (see also \cite{Falcioni:2019nxk} for recent work on relations between different soft functions), we have
\begin{align}
\gamma^r= -\cG_0+2B\,,
\end{align}
where $B$ is the virtual anomalous dimension (the coefficient of $\delta(1-x)$ in the DGLAP kernel), and $\cG_0$ is the collinear anomalous dimension. We then find
\begin{align}
  \label{eq:resformula_N4}
\frac{d\sigma^{(0)}}{dz} = &\; \frac{1}{4} \int\limits_0^\infty db\, b
  J_0(bQ\sqrt{1-z})H(Q,\mu_h) j^q_\EEC(b,b_0/b,Q) j^{\bar
  q}_\EEC(b,b_0/b,Q) S_\EEC( b,\mu_s, \nu_s) 
\nn\\
&\exp \left[ \gcusp \log^2\left( \frac{b^2 Q^2}{b_0^2} \right)  +2B\log\left( \frac{b^2 Q^2}{b_0^2} \right)  \right]\,.
\end{align}
Both the cusp anomalous dimension, as well as the $B$ anomalous dimension are known from integrability \cite{Eden:2006rx,Beisert:2006ez,Freyhult:2007pz,Freyhult:2009my,Fioravanti:2009xt}.  It is interesting that only the two anomalous dimensions that are known from integrability appear in the final result. The collinear anomalous dimension, which drops out of the final result in a conformal theory, is known to four loops in $\cN=4$ \cite{Dixon:2017nat}.

This formula describes the leading power asymptotics of the EEC in the $z\to 1$ limit to all orders in the coupling (Indeed, in $\cN=4$, it should also apply at finite coupling). The goal of this paper will be to start to understand the all orders structure of the subleading power corrections to this formula in $(1-z)$. While we do not have a complete factorization formula or all orders understanding, we will be able to deduce much of the structure from general consistency and symmetry arguments. Ultimately, we would like to be able to classify the operators that appear in the description of the subleading powers in this limit, and understand their renormalization group evolution. This paper represents a first step in this direction.

We conclude this section by noting that the result of \Eq{eq:resformula_N4} can also be derived in a conformal field theory by directly considering the structure of the four point correlator in the double light cone limit \cite{Korchemsky:2019nzm} (see also \cite{Alday:2013cwa}), and using the duality between the correlator and a Wilson loop \cite{Alday:2010zy}, as well as known results for the structure constants \cite{Eden:2012rr}. It would be interesting to understand systematically the OPE of the correlator in this limit and the operators that appear from this perspective. There has been some study of the double light cone limit in \cite{Alday:2015ota}. It would be interesting to understand it in more detail and develop a systematic OPE, much like exists in the collinear limit \cite{Alday:2010ku,Basso:2014jfa,Basso:2007wd,Basso:2010in,Basso:2015uxa,Basso:2013vsa,Basso:2013aha,Basso:2014koa,Basso:2014nra}. It would also be interesting if the recently introduced light ray OPE \cite{Kravchuk:2018htv,Kologlu:2019bco,Kologlu:2019mfz} can provide insight into this limit. However, we leave these directions to future work.

\section{Fixed Order Calculation of the EEC at Subleading Power}\label{sec:FO}

Having discussed the structure of the EEC in the back-to-back limit, as well as the factorization theorem describing its leading power asymptotics, in this section we begin our study of the subleading corrections in powers of $(1-z)$ by performing a fixed order calculation. This is important both for understanding the structure of the subleading power rapidity divergences for the EEC, and for providing the boundary conditions for the renormalization group studied later in the paper. We will perform this calculation both in QCD, as well as in $\cN=4$. We follow closely the calculation for the power corrections for $p_T$ presented in \cite{Ebert:2018gsn}. In \Sec{sec:intuition} we summarize some of the intuition derived from this calculation, which provides significant insight into the structure of the subleading power renormalization group evolution, which we will then study in more detail in \Sec{sec:RRG_NLP}.

\subsection{Leading Order Calculation in $\cN=4$ SYM and QCD}\label{sec:FO2}

In this section, we perform the leading order (LO) calculation of the EEC at NLP in both QCD and $\cN=4$. This section is rather technical, and assumes some familiarity with fixed order calculations at subleading power in SCET (see e.g. \cite{Moult:2016fqy,Moult:2017jsg,Ebert:2018lzn,Ebert:2018gsn} for more detailed discussions).

The EEC observable can be written as
\begin{align}
	\frac{\df \sigma_{\text{EEC}}}{\df  y} &= \sum_{a,b} \int \df \Phi_{V \to a+b+X} \left|\cM_{V \to a+b+X}\right|^2 \frac{E_a E_b}{q_V^2}\delta\left(y-\cos^2 \frac{\theta_{ab}}{2}\right) \,,
\end{align}
where we have used $y \equiv 1-z$, so that $y \to 0$ in the back-to-back limit. To perform the calculation it is convenient to write the observable definition in a boost invariant manner
\begin{align}
	\frac{\df \sigma_{\text{EEC}}}{\df  y} &= \frac{1}{2 (1-y) q_V^2}\sum_{a,b} \int \df \Phi_{V \to a+b+X} \left|\cM_{V \to a+b+X}\right|^2\, p_a\cdot p_b\,\delta\left[y - \left(1 -\frac{q_V^2p_a\cdot p_b}{2 p_a\cdot q_V p_b \cdot q_V}\right)\right] \,,
\end{align}
where $q_V^\mu$ is the momentum of the vector boson. This definition is convenient since if we are correlating a particular pair of particles $\{a,b\}$, we can boost the system such that the particles being correlated are back-to-back and the vector boson (or source) recoils against the unmeasured radiation in the perpendicular direction.

Given this setup, we begin by considering a single perturbative emission, which is sufficient for the LO calculation. We will denote by $p_a^\mu$ and $p^\mu_b$ the momenta of the particles we are correlating, and $k^\mu$ the momentum of the unmeasured radiation. This translates to the following choice of kinematics\footnote{Note that here $p_{a,b}^\mu$ and $k^\mu$ are outgoing while $q_V^\mu$ is incoming.}
\begin{align}\label{eq:kinematic}
	p_a^\mu &= (q^-_V - k^-) \frac{n^\mu}{2}\,,&& k^\mu = k^- \frac{n^\mu}{2} + k^+ \frac{\bn^\mu}{2} + k_\perp^\mu\,,\nn\\
	p_b^\mu &= (q^+_V - k^+) \frac{\bn^\mu}{2}\,,&& q_V^\mu = q_V^- \frac{n^\mu}{2} + q_V^+ \frac{\bn^\mu}{2} + k_\perp^\mu\,.
\end{align}
The measurement of the EEC observable then takes the form of the following constraint
\begin{align}
	y = 1 -  \frac{(q^-_V - k^-)(q^+_V - k^+) q_V^2}{q_V^+ (q^-_V - k^-) q_V^- (q^+_V - k^+)} = \frac{q^+_V q^-_V - q_V^2}{q^+_V q^-_V} = \frac{\kperp}{q_V^2 - \kperp}\,,
\end{align}
which we can rewrite  as a constraint on $\kperp$ 
\be
	\delta\left(y - \frac{\kperp}{q_V^2 - \kperp} \right) = \delta\left(\kperp - q_V^2\frac{y}{1-y}\right) \frac{q_V^2}{(1-y)^2}\,.
\ee
This extends the relation between the EEC and $p_T$ to subleading powers.

\subsection*{Master Formula}

Using this result for the measurement function, the cross section for the EEC with one emission reads
\begin{align}
	\frac{\df \sigma_{\text{EEC}}}{\df  y} &= \frac{1}{2 (1-y)^3}\sum_{a,b}\int \df \Phi |\cM_{V \to q\bar{q} g}|^2\, p_a \cdot p_b\,\delta\left(\kperp - q_V^2\frac{y}{1-y}\right)\nn\,.
\end{align}
For our particular choice of frame where $p_a^\mu$ has no $\perp$ component, we have \cite{Fleming:2007qr}
\begin{align}
	 \int \frac{\df^{d-2} p_{a\perp}}{(2\pi)^{d-2}} &=  \sum_n d^2 p_{\perp,n} \int \frac{\df^{d-2} p_{a\perp}}{(2\pi)^{d-2}}\delta^{(d-2)}(p^\mu_{a\perp}) = \frac{|\vec{p}_a|^{d-2}}{(2\pi)^{d-2}} \int \df \Omega_{d-2}\,.
\end{align}
Using
\begin{align}\label{eq:paPS}
	\int \dfslash^d p_a \delta_+(p_a^2) &=  \frac{1}{2} \int \frac{\df p_a^-}{(2\pi)} \frac{\df p_a^+}{(2\pi)} (2\pi) \theta(p_a^+ + p_a^-)\delta(p_a^+ p_a^- ) |\vec{p}_a|^2 \int  \frac{ \df\Omega_{d-2}}{(2\pi)^{d-2}} \nn\\
	&= \frac{1}{2(2\pi)^{d-1}}  \int \frac{\df p_a^-}{p_a^-} \theta(p_a^-) |p_a^-/2|^2\int \df \Omega_{d-2} = C \times \int \df p_a^- \,p_a^-\,,   
\end{align}
we can write the cross section with a single emission as
\begin{align}
	\frac{\df \sigma_{\text{\text{EEC}}}}{\df  y} &\sim \frac{1}{2 (1-y)^3}\sum_{a,b} \int \dfslash^d k \delta_+(k^2) |\cM_{V \to q\bar{q} g}|^2\, (p_a \cdot p_b)^2\,\delta\left(\kperp - q_V^2\frac{y}{1-y}\right)\nn\,\\
	&\sim \frac{1}{2 (1-y)^3}\sum_{a,b} \int_0^{q^-} \frac{\df k^-}{k^-}  \int \df \kperp \frac{\mu^{2\epsilon}}{\kperpeps} |\cM_{V \to q\bar{q} g}|^2\, (p_a \cdot p_b)^2\,\delta\left(\kperp - q_V^2\frac{y}{1-y}\right)\nn\,\\
	&\sim \frac{1}{2 (1-y)^3}\left(\frac{\mu^{2}}{q_V^2}\frac{1-y}{y}\right)^\epsilon \sum_{a,b} \int_0^{q^-} \frac{\df k^-}{k^-} |\cM_{V \to q\bar{q} g}|^2(k^-,y)\, (p_a \cdot p_b)^2 (k^-,y)\,,
\end{align}
where everything is a function of $k^-$, $y$ and $q_V^2$. Up to this point, we have not expanded anything, but we have enforced the measurement, momentum conservation and the choice of frame.  We can now express $k^-$ as a dimensionless fraction $x$ via
\be\label{eq:xdefinition}
	x = \frac{k^-}{q^-}\,,
\ee 
and we can write all Mandelstam invariants in terms of $x$ and $y$
\begin{align}
	k^-&= x q^- \,,\qquad \kperp = q_V^2\frac{y}{1-y}\,,\qquad k^+ = \frac{\kperp}{k^-} = \frac{q_V^2}{q^-}\frac{y}{1-y} \frac{1}{x}\,,\nn \\
	q^2_V &\equiv q^+ q^- - \kperp = q_V^+ q_V^- - q_V^2\frac{y}{1-y} \quad\implies\quad q_V^2 = q^+_V q^-_V(1-y)\,, \nn \\
	s_{ab}(x,y) &= 2p_a \cdot p_b = (q_V^- - k^-)(q_V^+ - k^+) = q_V^+q_V^- (1-x)\left(1 - \frac{y}{x}\right) \nn \\
	&= q_V^2 \frac{(1-x)}{(1-y)}\left(1 - \frac{y}{x}\right) \,,\nn\\
	s_{ak}(x,y) &= 2p_a \cdot k = (q_V^- - k^-)k^+ = q_V^2\frac{y}{(1-y)}\frac{(1-x)}{x}\,,\nn\\
	s_{bk}(x,y) &= 2p_b \cdot k = (q_V^+ - k^+)k^- = q_V^2 \frac{x}{(1-y)}\left(1 - \frac{y}{x}\right)\,.
\end{align}
With these expressions one can easily check that
\be
	s_{ab}+ s_{ak}+ s_{kb} = q_V^2\,,
\ee
with no power corrections.
We can now use the expressions for $s_{ab}$ to arrive at
\begin{align}
	\frac{\df \sigma_{\text{EEC}}}{\df  y} &= \frac{1}{(1-y)^5}\left(\frac{\mu^{2}}{q_V^2}\frac{1-y}{y}\right)^\epsilon \sum_{a,b} \int_0^{1} \frac{\df x}{x} (1-x)^2  \left(1 - \frac{y}{x}\right)^2|\cM_{V \to q\bar{q} g}|^2(x,y)\,.
\end{align}
Expanding in the collinear limit is now the same as expanding in $y$, and we find up to NLP
\begin{align}\label{eq:collinearMasterUnreg}
	\frac{\df \sigma_{\text{EEC}}}{\df  y} &=\left(\frac{\mu^{2}}{q_V^2 y}\right)^\epsilon \sum_{a,b} \int_0^{1} \frac{\df x}{x} (1-x)^2  \left[A^{(0)}(x)+A^{(2)}(x) + y A^{(0)}(x)\left( 5-\frac{2}{x} -\epsilon \right) \right]\,.
\end{align}
Here $A^{(0)}(x)$ and $A^{(2)}(x)$ are the expansion of the squared matrix elements in the collinear limits,
\begin{align}
|\cM_{V \to q\bar{q} g}|^2(x,y)=A^{(0)}(x)+A^{(2)}(x) + y A^{(0)}(x)+\cdots.
\end{align}

\subsection*{Rapidity Regulator}

The expression for the EEC in \eq{collinearMasterUnreg} is divergent as $x \to 0$. This is a rapidity divergence and must be regulated with a rapidity regulator. Here we present the result using pure rapidity regularization \cite{Ebert:2018gsn}, which greatly simplifies the calculation, particularly for the constant (the non-logarithmically enhanced term). We have also computed the logarithm using the more standard $\eta$-regulator \cite{Chiu:2012ir,Chiu:2011qc}, and find an identical result.

We take as a regulator the rapidity in the $n$-collinear sector, normalized by the rapidity of the color singlet\footnote{In the rest frame of the decaying boson this would be 1. Here we use a slightly boosted frame, so adding this factor is necessary to guarantee that the result is independent of the frame.}
\be
	e^{-Y_V}e^Y_n = \frac{q_V^+}{q_V^-}\frac{p_1^- + k^-}{p_1^+ + k^+} = \frac{q_V^+}{k^+} = \frac{q_V^+ q_V^- x}{\kperp} = \frac{q_V^+ q_V^- x (1-y)}{q_V^2 y } =  \frac{x}{y}\,.
\ee
The rapidity regulated result is then given by
\begin{align}\label{eq:collinearMasterReg}
	\frac{\df \sigma_{\text{EEC}}}{\df  y} &=\left(\frac{\mu^{2}}{q_V^2 y}\right)^\epsilon \frac{y^\eta}{\upsilon^\eta} \sum_{a,b} \int_0^{1} \frac{\df x}{x^{1+\eta}} (1-x)^2  \left[A^{(2)}(x) + y A^{(0)}(x)\left( 5-\frac{2}{x} -\epsilon \right) \right]\,,
\end{align}
which can be straightforwardly integrated.

\subsection*{Results}

Plugging in the expression for $A^{(0)}$ and $A^{(2)}$ in QCD and $\cN=4$ gives the result for the EEC  up to NLP
\begin{align}\label{eq:results}
	\frac{1}{\sigma_0}\frac{\df \sigma_{\text{EEC}}^\text{QCD}}{\df  y} &= -\frac{1}{y}\left(\log y + \frac{3}{4}\right) - 5\log y-\frac{9}{4} +\cO(y)\,, \nn \\
	\frac{1}{\sigma_0}\frac{\df \sigma_{\text{EEC}}^{\cN=4}}{\df  y} &=- \frac{\log y}{y}- 2\log y +\cO(y)\,.
\end{align}
This agrees with the expansion of the full angle result for $\cN=4$ in \cite{Belitsky:2013ofa,Belitsky:2013xxa,Belitsky:2013bja}, as well as the classic QCD result \cite{Basham:1978bw} for both the logarithm and the constant. This agreement (in particular for the constant) illustrates that we have the correct effective field theory setup, and that we understand the regularization of rapidity divergences at subleading power. This is further supported by our calculation of the subleading power corrections for the case of $p_T$ in \cite{Ebert:2018gsn}, which was performed using the same formalism. While this expansion is an inefficient way to compute these subleading terms, which can much more easily be obtained by performing the full calculation and expanding the result, the ability to systematically compute the terms at each order in the power expansion will allow us to perform an all orders resummation by deriving renormalization group evolution equations in rapidity.

\subsection{Physical Intuition for Subleading Power Rapidity Divergences}\label{sec:intuition}

In this section we wish to summarize some of the general lessons learned from the above fixed order calculation (as well as from the calculation of the fixed order subleading rapidity logarithms for $p_T$ in \cite{Ebert:2018gsn}), which provides significant clues into the structure of the subleading power rapidity renormalization group.  Since we have shown above that the description of the EEC in the back-to-back limit is ultimately formulated in the EFT in terms of transverse momentum, here we will phrase all the functions in terms of transverse momentum, however, these can straightforwardly be converted back to derive results for the EEC.
We will also phrase this discussion in terms of the $\eta$ regulator \cite{Chiu:2012ir,Chiu:2011qc}, due to the fact that it is more familiar for most readers.

The first important observation from the fixed order calculation is that at NLP there are no purely virtual corrections at lowest order in $\alpha_s$ (such a correction would appear as $y\delta(y)=0$). The lowest order result for the EEC in $\cN=4$, which we use as an example due to its simpler structure, is given by
\begin{align}\label{eq:N4_NLP_fixedorder}
	\frac{1}{\sigma_0}\frac{\df \sigma_{\text{EEC}}^{\cN=4}}{\df  (1-z)} &=  - 2\log(1-z) +\cO(1-z)\,.
\end{align}
Here the single logarithm comes from real soft or collinear emissions. Since the soft and collinear sectors lie at the same virtuality, this guarantees that at subleading power the lowest order logarithm must be a rapidity logarithm. This can be made explicit by writing down a general ansatz for the one-loop result. This approach was first introduced in the \sceti~ case in \cite{Moult:2016fqy}.  Here we will phrase it in terms of the underlying $p_T$ dependence, since this is perhaps more familiar to the resummation community. The general form of the one-loop result for the NLP corrections to $p_T$ can be written as
\begin{align}
\frac{d\sigma^{(2)}}{dp_T^2}=\left( \frac{\mu^2}{p_T^2}  \right)^\epsilon   \left( \frac{\nu}{p_T}  \right)^\eta  \left[  \frac{s_\epsilon}{\epsilon}+\frac{s_\eta}{\eta}  \right] +\left( \frac{\mu^2}{p_T^2}  \right)^\epsilon   \left( \frac{\nu}{Q}  \right)^\eta  \left[  \frac{c_\epsilon}{\epsilon}+\frac{c_\eta}{\eta}  \right] \,.
\end{align}
Expanding this, we find
\begin{align}
\frac{d\sigma^{(2)}}{dp_T^2}=\left( \frac{c_\epsilon+s_\epsilon}{\epsilon}  \right)   +     \left( \frac{c_\eta+s_\eta}{\eta}  \right) +2(c_\epsilon+s_\epsilon)  \log\frac{\mu}{p_T} +s_\eta \log \frac{\nu}{p_T} +c_\eta \log\frac{\nu}{Q}\,.
\end{align}
Demanding that there are no $1/\epsilon$ or $1/\eta$ poles in the final answer imposes the conditions 
\begin{align}
c_\epsilon=-s_\epsilon\,, \qquad c_\eta=-s_\eta\,,
\end{align}
and shows that the lowest order logarithm appearing in the cross section is a rapidity logarithm.

This simple observation provides considerable insight into the structure of the subleading power renormalization group evolution equations. In \cite{Moult:2018jjd} it was shown that the way that a single logarithm at the first non-vanishing order can be generated is through renormalization group mixing. In the particular case studied in \cite{Moult:2018jjd} there was only a virtuality renormalization group, and therefore the single logarithm was generated by the $\mu$-RGE. However, here we see that for observables that have both a $\mu$ and $\nu$ RGE, this mixing will always occur in the $\nu$ RGE, since the lowest order logarithm is always a rapidity logarithm. 

Our fixed order calculation also provides insight into the structure of the cancellation of rapidity divergences at subleading power.  Although we have not written down a complete set of \scetii~operators, we briefly comment on the physical intuition for the cancellation of rapidity anomalous dimensions, which will then determine how renormalization group consistency appears in the effective theory. We can consider the case of the NLP correction from a soft quark, since this will provide the clearest picture of the differences between rapidity divergences and virtuality divergences. At lowest order, the soft function for the emission of a soft quark is given by
\begin{align}\label{eq:quark_soft}
S^{(2)}_q&= \fd{3cm}{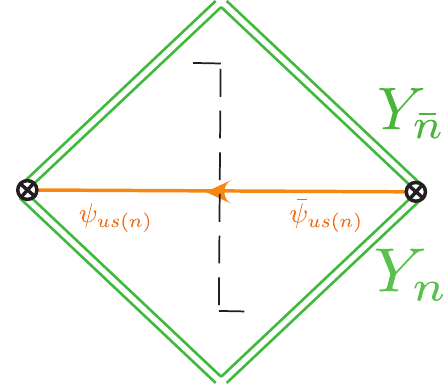}\nn \\
&\propto \mu^{2\epsilon} \nu^\eta \int \dbar^d k |k^+-k^-|^{-\eta} \delta^+ (k^2) \frac{(p_\perp^2)^{-\epsilon}}{\Gamma(1-\epsilon) \pi^\epsilon} \delta^{d-2} (p_\perp-\hat \cP_\perp) \\
&=\frac{1}{\eta} + \log{\frac{\nu}{p_\perp}} +\cdots\,.
\end{align}
This soft function is $\epsilon$ finite, but exhibits the expected $\eta$ divergence. This divergence cannot be absorbed into the renormalization of this soft function, since it starts at $\alpha_s$, and therefore we must find the class of operators that this soft function mixes into. We will derive the structure of these operators in \Sec{sec:RRG_NLP}.

While the fact that this operator must mix into a new operator is familiar from other studies at NLP, what is different is that the integrand for the soft function is ``1".  This implies that an equal part of the divergence comes from when the quark goes to infinite rapidity in either direction. This has an interesting interpretation, which can guide the physical intuition for how the cancellation of rapidity divergences occurs. As the soft quark goes collinear in the direction of the  collinear quark, the rapidity divergence must be cancelled by a collinear rapidity divergence from a subleading power jet function describing two collinear quarks. One the other hand, as the soft quark goes collinear with the gluon, it must be cancelled by a subleading power jet function involving a collinear quark and gluon field. In pictures, the two limits are shown as
\begin{align}
\fd{3cm}{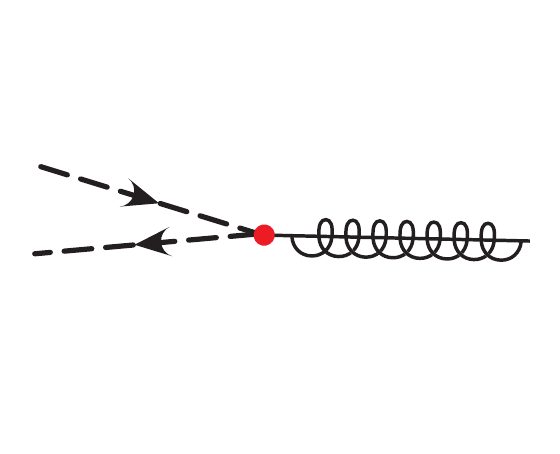}\xleftarrow[]{\eta\to -\infty} \fd{3cm}{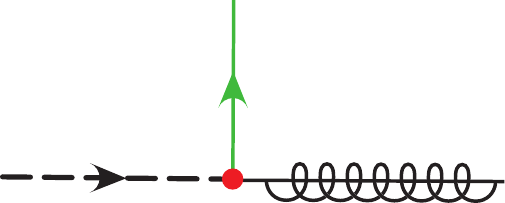}  \xrightarrow[]{\eta\to \infty} \fd{3cm}{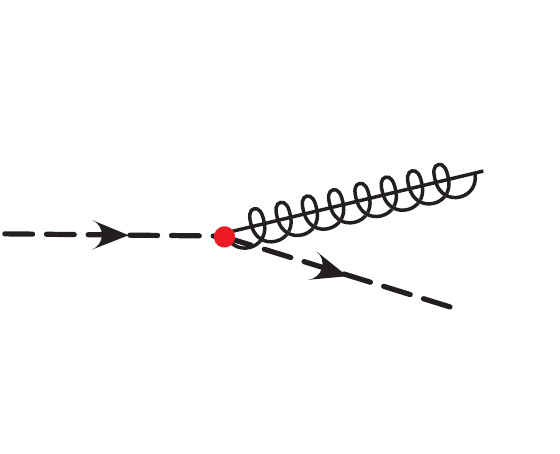}\,.
\end{align}
This is of course exactly what happens at leading power, however, at leading power the jet functions are identical in all limits, giving a much simpler structure. The structure for the cancellation of the rapidity divergences at subleading power implies that renormalization group consistent subsets of operators will appear in triplets, as opposed to doublets, as was observed in \cite{Moult:2018jjd} for the $\mu$-RGE. The factorization formula for a single pair of triplets will take the (extremely) schematic form
\begin{align}
\frac{d\sigma^{(2)}}{dz}&=H_1 J_{\bar n}^{(2)} J_n^{(0)} S^{(0)} 
+H_2 J_{\bar n}^{(0)} J_n^{(2)} S^{(0)}
+H_3 J_{\bar n}^{(0)} J_n^{(0)} S^{(2)}\,,
\end{align}
where the jet and soft functions with superscript $(2)$ denote power suppressed functions. In terms of pictures, we have
\begin{align}\label{eq:schematic_factorization}
\frac{d\sigma^{(2)}}{dz}&=\nn \\
&\underbrace{  \left| \fd{1.5cm}{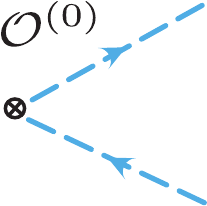}~ \right|^2  \cdot   \int dr_2^+ dr_3^+ \fd{3cm}{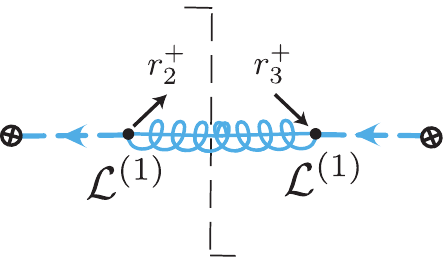}\otimes \fd{3cm}{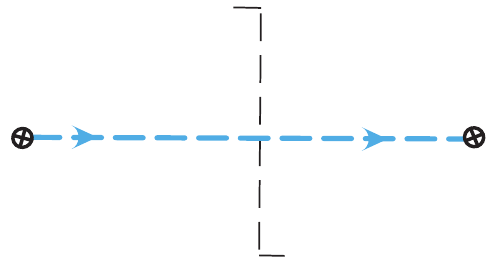} \otimes  \fd{3cm}{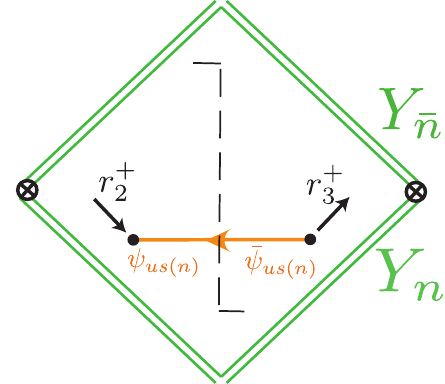}   }_{\text{Soft Quark Correction}}\nn \\
 +&\underbrace{ \int d \omega_1 d \omega_2 \left| \fd{1.5cm}{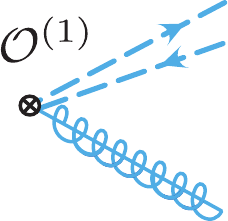}~ \right|^2\otimes\fd{3cm}{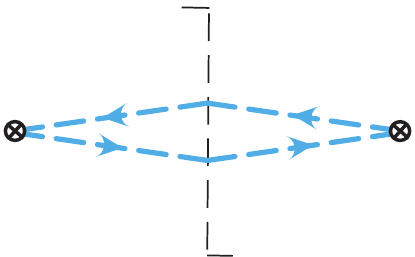}\otimes \fd{3cm}{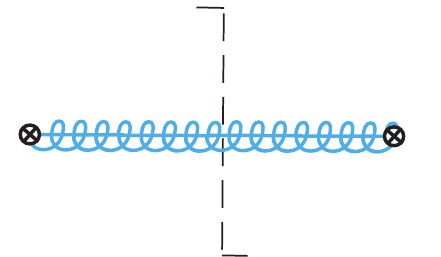}\otimes \fd{3cm}{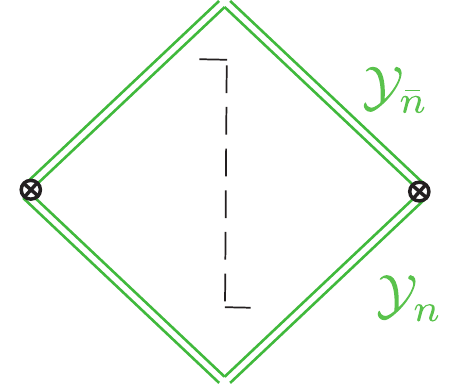} }_{\text{Collinear Quark Correction 1}} \nn \\
 +&\underbrace{ \int d \omega_1 d \omega_2 \left| \fd{1.5cm}{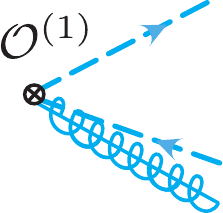}~\right|^2  \otimes~\fd{3cm}{figures/1quark_jetfunction_low}\otimes  \fd{3cm}{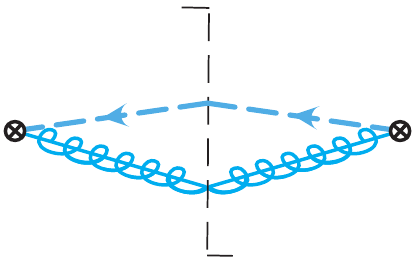}\otimes \fd{3cm}{figures/eikonal_factor_gluon_low.pdf}}_{\text{Collinear Quark Correction 2}}\,,
\end{align}
which perhaps makes more clear schematically how the cancellation between $\nu$ anomalous dimension occurs between the soft and collinear sectors of the theory. A similar triplet exists from the corrections associated with soft gluon emission. Each of the power suppressed functions in \Eq{eq:schematic_factorization} will mix in rapidity, as was illustrated for the soft function in \Eq{eq:quark_soft}. To perform the renormalizaton, we must therefore identify which operators are being mixed into in each case.

It is interesting to compare this to the structure of the subleading power factorization for \sceti~ event shapes, which is described in \cite{Moult:2018jjd,Moult:2019uhz}. There subleading power jet functions involving two quark fields, and those involving a quark and gluon field are in separately renormalization group consistent pairings. Therefore, we find that the \scetii~ case gives rise to a much tighter structure in the EFT, since it links multiple hard scattering operators. Note that this also suggests that the issue of endpoint divergences is harder to avoid, although this is a topic that we leave for future study.

In conclusion, we have learned a number of general lessons about the subleading power $\nu$-RG from our perturbative calculation. In particular, we have seen that subleading power corrections to the EEC (and more generally any observable that involves both rapidity and virtuality renormalization group flows) involve a mixing in rapidity at the first non-trivial order into a new class of operators. In \Sec{sec:RRG_NLP} we will derive the structure of these new operators by studying the consistency of the structure of the renormalization group.

\section{Rapidity Renormalization Group at Subleading Power}\label{sec:RRG_NLP}

In the previous section we have shown that for the EEC, and more generally for subleading corrections to observables with both rapidity and virtuality scales, the subleading power rapidity RGE will always involve mixing into additional operators. The goal of this section will be to derive the structure of the operators that arise from this mixing, as well as their renormalization group properties. As in our study of thrust at subleading powers \cite{Moult:2018jjd}, our approach to gain a handle on the structure of the RG equations at subleading power will be to use an illustrative example of subleading power jet and soft functions whose renormalization group structure can be obtained from the known leading power RG equations. Once the particular form of the operators is derived, they can then be applied in other situations. 

We will find that for the case of rapidity divergences considered here, the use of an illustrative example is more subtle than for the $\mu$ RGE due to the appearance of additional divergences that appear. We will also find that due to this, there are multiple (two distinct) operators that can be mixed into. Therefore, our analysis in \Sec{sec:example} should be viewed as providing motivation for the type of operators that appear, although in the initial form that they arise, they will involve unregularized integrals. With this motivation for their structure, we are then able to use the commutativity of the $\mu$ and $\nu$ RGEs in \Sec{sec:consistency} to fix the RG properties of these operators and provide regularized definitions. We then solve the associated evolution equations for the newly introduced operators in \Sec{sec:solution}.

\subsection{An Illustrative Example}\label{sec:example}

We will begin by considering the LP soft function for $p_T$, defined as
\begin{align}
 S(\vec p_T) &= \frac{1}{N_c^2 - 1} \big\langle 0 \big| \mathrm{Tr} \bigl\{
   \mathrm{T} \big[\cS^\dagger_{\bn} \cS_n\big]
  \delta^{(2)}(\vec p_T-\cP_\perp) \overline{\mathrm{T}} \big[\cS^\dagger_{n} \cS_{\bn} \big]
   \bigr\} \big| 0 \big\rangle
\,.\end{align}
This soft function satisfies the $\mu$ and $\nu$ RGEs
\begin{align}
\nu \frac{d}{d\nu}S(\vec p_T)&=\int d\vec q_T \gamma^S_\nu(p_T-q_T) S(\vec q_T)\,, \nn \\
\mu \frac{d}{d\mu}S(\vec p_T)&=\gamma_\mu^S S(\vec p_T)\,,
\end{align}
with the anomalous dimensions
\begin{align}
\gamma_\mu^S &=4\gcusp(\alpha_s)\log \left( \frac{\mu}{\nu}\right)\,,\nn \\
\gamma_\nu^S &=2\gcusp (\alpha_s)\cL_0\left( \vec p_T,\mu \right)\,.
\end{align}
Crucially, the $\mu$ anomalous dimension is multiplicative, while the $\nu$ anomalous dimension is a convolution in $p_T$ space. It is this fact that will ultimately lead to the subleading power rapidity renormalization group having a more interesting structure. 

A simple trick to understand the structure of subleading power RG equations that was first used in \cite{Moult:2018jjd} is to consider jet and soft functions obtained by multiplying the leading power jet and soft functions by a kinematic invariant. In the present case, we 
can consider the subleading power soft function defined by
\begin{align}
S_{p_T^2}^{(2)}(\vec p_T)=\vec p_T^2 S(\vec p_T)\,.
\end{align}
The superscript $(2)$ indicates that this function is power suppressed due to the explicit factor of $p_T^2$, and the subscript $p_T^2$ is meant to identify the nature of this power suppression. The structure of this function is known to all orders, since it is inherited from the known structure of the leading power soft function. However, understanding how this structure is manifested in the renormalization group structure of $S_{p_T^2}^{(2)}$ is non-trivial, and will reveal new operators.

First, we note that the $\mu$ RGE for this subleading power soft function is identical to the RGE of the leading power soft function, since the $\mu$ RGE is multiplicative. We therefore have
\begin{align}
\mu \frac{d}{d\mu}S_{p_T^2}^{(2)}(\vec p_T)=\gamma^\mu_S   S_{p_T^2}^{(2)}(\vec p_T)\,.
\end{align}
However, we find a more interesting behavior for the $\nu$ RGE, due to the fact that it is a convolution in $p_T$. Multiplying both sides of the LP RGE by $\vec p_T^2$, and using the identity
\begin{align}\label{eq:pt_expand}
\vec p_T^2 =(\vec p_T-\vec q_T)^2 +\vec q_T^2 +2(\vec p_T- \vec q_T) \cdot \vec q_T\,,
\end{align}
we arrive at the equation
\begin{align}
\nu \frac{d}{d\nu}S_{p_T^2}^{(2)}(\vec p_T)&=\int d\vec q_T  (\vec p_T-\vec q_T)^2  \gamma_S(p_T-q_T) S(q_T)  \\
&+\int d\vec q_T  \gamma_S(p_T-q_T) \left[ 2(\vec p_T- \vec q_T) \cdot \vec q_T S(q_T)\right] + \int d\vec q_T \gamma_S(p_T-q_T) \left[ \vec q_T^2 S(\vec q_T)\right]\,. \nn
\end{align}
Note that we must arrange \Eq{eq:pt_expand} in this form, so that it is a kernel in $\vec p_T- \vec q_T$ multiplying a function of $q_T$.
Simplifying this result, we find that we can write it  as
\begin{align}
\nu \frac{d}{d\nu}S_{p_T^2}^{(2)}(\vec p_T)&=2\gcusp {\mathbb{I}}_S \\
&+\int d\vec q_T  \gamma_S(p_T-q_T) 2(\vec p_T- \vec q_T) \cdot   \vec S^{(1)}(q_T) +\int d\vec q_T \gamma_S(p_T-q_T) S_{p_T^2}^{(2)}(\vec q_T)\,.\nn
\end{align}
Here we see a renormalization group mixing with two power suppressed functions. The first is 
\begin{align}
{\mathbb{I}}_S\propto\int d^2\vec q_T~   S(\vec q_T)\,,
\end{align}
which we will refer to as the ``rapidity identity operator", and will play a crucial role in our subsequent analysis. As written, the integral over $q_T$ goes to infinity, and therefore this expression is ill-defined, and will require regularization. For this reason we have also been glib about what this operator depends on. In the next section we will present a way of deriving its renormalization group properties, as well as a regularized definition. The goal of this section is merely to illustrate that the subleading power soft function mixes into a new operator which is loosely related to a moment of the leading power soft function. The second operator arising in the mixing is 
\begin{align}
\vec S^{(1)}(\vec p_T)=\vec p_T S^{(0)}(\vec p_T)\,,
\end{align}
which is a vector soft function which scales like $\cO(\lambda)$. This function can only appear in a factorization formula if it is dotted into a vector jet function, or some other vector quantity.

While we believe that it would be extremely interesting to study in more detail the complete structure of this illustrative example, and we will return to this in future work, here we focus only on the elements of these equations that are required at leading logarithmic accuracy. We note that
\begin{align}
{\mathbb{I}}_S=1+\cO(\alpha_s)\,, \qquad S^{(1)}(\vec p_T)=0+\cO(\alpha_s)\,.
\end{align}
Therefore, in the leading logarithmic series, $S_{p_T^2}^{(2)}$ mixes into ${\mathbb{I}}_S$, and we can ignore $S^{(1)}$. We can therefore simplify our equation to
\begin{align}
\nu \frac{d}{d\nu}S_{p_T^2}^{(2)}(\vec p_T,\mu,\nu)&=2\gcusp {\mathbb{I}}_S  +\int d\vec q_T \gamma_S(p_T-q_T) S_{p_T^2}^{(2)}(\vec q_T,\mu,\nu)\,.
\end{align}
We see that what is occurring is that the power suppressed soft function is mixing with the rapidity identity operator. This provides a renormalization group derivation of the perturbative calculations in \Sec{sec:FO}, where one generates a rapidity divergence and associated logarithm at the lowest non-trivial order. This is simply a perturbative description of the mixing into ${\mathbb{I}}_S$. Now, with this general structure in mind, we would like to understand the properties of this rapidity identity operator.

We emphasize again that we have performed only a cursory study of this illustrative example so as to be able to illustrate the renormalization group mixing in rapidity, and to identify the structure of the rapidity identity operator. It would be particularly interesting to study the complete structure of this illustrative example to all logarithmic orders, and in particular to better understand the structure of the convolutions in $p_T$. However, since the focus of this paper is on deriving the leading logarithmic series for the EEC, we will leave this to future work.

\subsection{Rapidity Identity Operators}\label{sec:consistency}

In the previous section, we found that the subleading power rapidity renormalization group involves a mixing into the rapidity identity operator, which is loosely related to the first moment of the leading power soft function
\begin{align}\label{eq:identity_unregularized}
{\mathbb{I}}_S\propto\int d^2\vec q_T~   S(\vec q_T)\,.
\end{align}
The goal of this section will be to understand how to make sense of this operator, since it is ill-defined as currently written. This is a crucial difference as compared with the case of thrust considered in \cite{Moult:2018jjd}. There a similar first moment operator appears, defined as \cite{Moult:2018jjd}
\begin{align}\label{eq:theta_soft_first}
S^{(2)}_{g,\theta}(k,\mu)&= \frac{1}{(N_c^2-1)} \tr \langle 0 | \cY^T_{\bar n} (0)\cY_n(0) \theta(k-\hat \Tau) \cY_n^T(0) \cY_{\bar n}(0) |0\rangle\,.
\end{align}
However, there the first moment is a finite integral, and does not introduce new divergences, allowing all the properties of this operator to be immediately deduced from those of the leading power operator. For the case of $p_T$ considered here, additional arguments must be used to fully fix the structure of the rapidity identity operator.

The $\mu^2$ dependence of the rapidity identity operator can be derived using the commutativity of the RG \cite{Chiu:2012ir,Chiu:2011qc}, namely that
\begin{align}
\left[  \frac{d}{d\mu}, \frac{d}{d\nu}  \right]=0\,,
\end{align}
which ensures the path independence of the $\mu$ and $\nu$ rapidity evolution. Here we consider a general subleading power soft operator $S^{(2)}$ (not necessarily $S_{p_T^2}^{(2)}$). If we assume that this operator mixes into a single identity type operator, then we obtain
\begin{align}
\mu \frac{d}{d\mu} \left[  \nu \frac{d}{d\nu} S^{(2)}  \right]&=\mu \frac{d}{d\mu} \left[ \gamma_{\delta{\mathbb{I}} }  {\mathbb{I}}_S+\gamma_\nu S^{(2)}  \right]\,,\nn \\
\nu \frac{d}{d\nu} \left[  \mu \frac{d}{d\mu} S^{(2)}  \right]&= \nu\frac{d}{d\nu} \left[  \gamma^S_\mu S^{(2)} \right]\,.
\end{align}
Here, we have used $\gamma_{\delta{\mathbb{I}} }$ to denote the mixing anomalous dimension.
Performing the next differentiation, we  then obtain the equality
\begin{align}
\gamma_{\delta{\mathbb{I}} } \mu \frac{d}{d\mu} {\mathbb{I}}_S +\left[  \mu \frac{d}{d\mu} \gamma_\nu \right]S^{(2)} +\gamma_\nu \gamma_\mu^S S^{(2)} =\gamma_\mu^S \gamma_{\delta{\mathbb{I}} }  {\mathbb{I}}_S +\left[ \nu \frac{d}{d\nu} \right]S^{(2)} +\gamma_\nu \gamma_\mu^S S^{(2)}\,.
\end{align}
Using the fact that commutativity is satisfied for the leading power anomalous dimensions, we arrive at
\begin{align}
\mu \frac{d}{d\mu}{\mathbb{I}}_S = \gamma_\mu^S {\mathbb{I}}_S\,.
\end{align}
This fixes the $\mu$ anomalous dimension of the rapidity identity operator.

To fix the $\nu$ anomalous dimension, we now apply commutativity to ${\mathbb{I}}_S$ itself, and use the fact that we know the $\mu$ anomalous dimension. We then have
\begin{align}
\left[  \frac{d}{d\mu}, \frac{d}{d\nu}  \right]  {\mathbb{I}}_S =0\,,
\end{align}
which gives the equation (at lowest order in $\alpha_s$, which is sufficient for LL)
\begin{align}
\mu \frac{d}{d\mu} \left( \nu \frac{d}{d\nu}  \right) {\mathbb{I}}_S=-4 \gcusp\,,
\end{align}
which can then be solved for
\begin{align}
\nu \frac{d}{d\nu} {\mathbb{I}}_S= -4 \gcusp \log \left( \frac{\mu^2}{\Lambda^2}  \right)\,,
\end{align}
where $\Lambda^2$ is an as yet to be determined scale. The only available scales at leading logarithmic accuracy are $\Lambda^2=p_T^2, \mu^2, \nu^2$. The cases $\Lambda^2=p_T^2, \nu^2$ both give the same behavior for the $\nu$ RGE on the hyperbola $\mu=p_T$, and therefore we will not treat them separately. It would be interesting to explore in more detail  the differences between these RGEs.

We therefore find two distinct identity operators with two different rapidity anomalous dimensions
\begin{align}
\nu \frac{d}{d\nu}{\mathbb{I}}^\nu_S(\mu, \nu) &= -\gamma_\mu^S {\mathbb{I}}^\nu_S(\mu, \nu)\,, \nn \\
\nu \frac{d}{d\nu}{\mathbb{I}}^{p_T^2}_S(p_T^2,\mu, \nu) &= 2 \gcusp \log \left( \frac{p_T^2}{\mu^2}  \right) {\mathbb{I}}^{p_T^2}_S(p_T^2,\mu, \nu)\,.
\end{align}
Here we have again used the superscript $p_T^2$ to indicate the identity function that the $S_{p_T^2}^{(2)}$ function mixes into, as we will argue shortly, and the superscript $\nu$ to indicate the identity function that has a non-trivial $\nu$ RGE on the $\mu=p_T$ hyperbola.

While this argument allows us to derive the renormalization group properties of these operators, which is sufficient for the purposes of this paper, it is also interesting to give explicit example of functions that realize this behavior. This is easy to do by defining regularizations of the integral in \Eq{eq:identity_unregularized}. 
Functions which give the behavior of the different identity operators at LL are
\begin{align}
{\mathbb{I}}^\nu_S(\mu, \nu)&=\int_0^{\nu^2} d^2\vec q_T~   S(\vec q_T)\,, \\
{\mathbb{I}}_S^{p_T^2}(p_T^2,\mu, \nu)&=\int_0^{p_T^2} d^2\vec q_T~   S(\vec q_T)\,.
\end{align}
The first function is a function of only $\mu^2/\nu^2$, while the second also depends on $p_T^2$. These two functions have different properties under boosts, and therefore do not themselves mix. This provides leading logarithmic definitions of the rapidity identity operators. It would be extremely interesting to understand how to extend these definitions beyond leading logarithm, however, we leave this to future work.

For the particular soft function considered in our illustrative example, $S_{p_T^2}^{(2)}=p_T^2 S^{(0)}$, one can use the knowledge of the two loop soft function to show that it is the operator ${\mathbb{I}}^{p_T^2}_S(p_T^2,\mu, \nu)$ that is being mixed into. This can also be argued directly by symmetry grounds: at the scale $\mu=p_T$, the leading power soft function does not flow in $\nu$ at leading logarithmic accuracy. This is due to its boost invariance. This property is not broken by multiplying by $p_T^2$, and therefore must also be a property of the counterterm operator that is being mixed with. This identifies the operator ${\mathbb{I}}^{p_T^2}_S(p_T^2,\mu, \nu)$. We can therefore make more precise the equation earlier for the RG of this function
\begin{align}
\nu \frac{d}{d\nu}S_{p_T^2}^{(2)}(\vec p_T,\mu,\nu)&=2\gcusp {\mathbb{I}}^{p_T^2}_S(\vec p_T,\mu,\nu)  +\int d\vec q_T \gamma_S(p_T-q_T) S_{p_T^2}^{(2)}(\vec q_T,\mu,\nu)\,.
\end{align}
For subleading power soft functions with explicit fields inserted, this argument no longer holds, and one can mix into the other operator.

It is also interesting to arrive at these conclusions for the structure of the renormalization group by manipulation of the renormalization group equations. This approach is ultimately ill-defined due to the lack of convergence of the integrals, but it gives the same results as derived from the commutativity of the RG, and provides additional insight into the origin of this behavior.
Recall that the leading power renormalization group evolution equations are
\begin{align}
\nu \frac{d}{d\nu}S(\vec p_T)&=\int d\vec q_T \gamma^S_\nu(p_T-q_T) S(q_T)\,, \nn \\
\mu \frac{d}{d\mu}S(\vec p_T)&=\gamma_\mu^S S(\vec p_T)\,,
\end{align}
with the anomalous dimensions
\begin{align}
\gamma_\mu^S &=4\gcusp(\alpha_s)\log \left( \frac{\mu}{\nu}\right)\,,\nn \\
\gamma_\nu^S &=2\gcusp (\alpha_s)\cL_0\left( \vec p_T,\mu \right)\,.
\end{align}
Since the $\mu$ anomalous dimension is multiplicative, this should not be changed if we integrate over $q_T$. In other words, we have
\begin{align}
\int d^2 p_T \left[ \mu \frac{d}{d\mu}S(\vec p_T)=\gamma^\mu_S S(\vec p_T) \right] \implies \mu \frac{d}{d\mu} {\mathbb{I}}_S=  \gamma^\mu_S {\mathbb{I}}_S\,,
\end{align}
which immediately leads to the fact that ${\mathbb{I}}_S$ is multiplicatively renormalized in $\mu$ with the same anomalous dimension as the leading power soft function.  For the $\nu$ renormalization group equation, we have
\begin{align}
\nu \frac{d}{d\nu} {\mathbb{I}}_S &=\int d^2 \vec q_T \nu \frac{d}{d\nu} S(\vec q_T)\nn \\
&=\int d^2 \vec q_T \left[   \int d^2 \vec p_T \gamma_S(\vec q_T-\vec p_T) S(\vec p_T)  \right]\,.
\end{align}
Now, performing the shift $\vec q_T\to \vec q_T +\vec p_T$, we obtain
\begin{align}
\nu \frac{d}{d\nu} {\mathbb{I}}_S &=\left[ \int d^2 \vec q_T \gamma_S(\vec q_T)  \right]{\mathbb{I}}_S\,,
\end{align}
which is again multiplicative. The expression in square brackets is not well defined, and must be fixed by some regularization, as was shown above. In this case, this shift argument may no longer be valid. However, this exercise is merely meant to illustrate another perspective on why the rapidity identity operator should satisfy a multiplicative $\nu$ RGE, and the argument presented earlier in this section should be taken as primary.

Therefore, in summary, we have shown that there are non-trivial rapidity identity operators that arise at subleading power, and we have identified the renormalization group properties of these operators at leading logarithmic accuracy. The first rapidity identity operator does not depend on the observable, and its anomalous dimensions are given by
\begin{align}\label{eq:rap_identity_summary}
\mu \frac{d}{d\mu} {\mathbb{I}}^\nu_S \left(  \frac{\mu^2}{\nu^2} \right) &= -2\gcusp\log\left(  \frac{\nu^2}{\mu^2} \right)~ {\mathbb{I}}^\nu_S\left(  \frac{\mu^2}{\nu^2} \right)\,,  \nn \\
\nu \frac{d}{d\nu} {\mathbb{I}}^\nu_S\left(  \frac{\mu^2}{\nu^2} \right) &= 2\gcusp \log\left(  \frac{\nu^2}{\mu^2} \right)~ {\mathbb{I}}^\nu_S\left(  \frac{\mu^2}{\nu^2} \right)\,.
\end{align}
The second rapidity identity operator depends on the observable, and its anomalous dimensions are given by
\begin{align}\label{eq:rap_identity2_summary}
\mu \frac{d}{d\mu} {\mathbb{I}}^{p_T^2}_S \left( p_T^2,\mu, \nu \right) &= -2\gcusp\log\left(  \frac{\nu^2}{\mu^2} \right)~ {\mathbb{I}}^{p_T^2}_S\left( p_T^2,\mu, \nu \right)\,, \nn \\
\nu \frac{d}{d\nu} {\mathbb{I}}^{p_T^2}_S\left( p_T^2,\mu, \nu \right) &= 2\gcusp\log\left(  \frac{p_T^2}{\mu^2} \right)~ {\mathbb{I}}^{p_T^2}_S\left( p_T^2,\mu, \nu \right)\,.
\end{align}
Here we have written the anomalous dimension in terms of the cusp anomalous dimension \cite{Korchemsky:1987wg}, which is given by $\gcusp=4(\alpha_s/4\pi)C_A+\cO(\alpha_s^2)$.
We expect these functions to appear ubiquitously in subleading power rapidity renormalization, and we therefore believe their identification is an important first step towards an understanding of the structure of subleading power rapidity logarithms.

One also has rapidity identity operators in the jet/beam sectors, that are defined analogously to the soft operators. Their anomalous dimensions are fixed to be identical to the soft rapidity identity operators (up to a sign) by RG consistency. For the particular case of the EEC, we can avoid them by always running to the jet scale. They are interesting for the case of $p_T$ resummation where one must consider their matching onto the parton distribution functions (PDFs). We will present a more detail discussion of these structures in future work.

It is also interesting to note that the subleading power Regge limit for massive scattering amplitudes has recently been studied in $\cN=4$ SYM in \cite{Bruser:2018jnc}. Their solution also involves an interesting operator mixing. Since there are connections between the Regge limit and the EEC at leading power due to conformal transformations, it would be interesting to understand if these persist at subleading power.

\subsection{Analytic Solution of Renormalization Group Evolution Equations}\label{sec:solution}

In this section we provide an analytic solution to the renormalization group evolution equations of the subleading power soft functions when mixing with either type of identity operator. Here we will consider only the case of fixed coupling, since our current application will be to $\cN=4$ (which is conformal), and the reader who is interested in extending these results to running coupling can consult  \cite{Moult:2018jjd}. We will also only study the renormalization group flow in $\nu$ at the scale $\mu=p_T$ to simplify the analysis. This will be sufficient for our applications, since we can always run the hard function down to the scale $\mu=p_T$. We will consider separately the two different types of identity operators, since we know that due to their different properties under boosts, they themselves cannot mix to generate a more complicated RG structure.

\subsection*{Identity Operator ${\mathbb{I}}^{p_T^2}_S \left( p_T^2,\mu, \nu \right)$:}

We first consider the case of mixing with the operator ${\mathbb{I}}_S \left( p_T^2,\mu, \nu \right)$, which gives rise to a simple $\nu$ RGE at the scale $\mu=p_T$. In this case, we get the RGE
\begin{align}
\nu \frac{d}{d\nu}\left(\begin{array}{c} S^{(2)} \\  {\mathbb{I}}^{p_T^2}_S \end{array} \right) &=  \left( \begin{array}{cc}0&\gamma_{\delta{\mathbb{I}} }\,  \\  0 & 0   \end{array} \right) \left(\begin{array}{c} S^{(2)} \\  {\mathbb{I}}^{p_T^2}_S \end{array} \right) \,, 
\end{align}
with the boundary conditions
\begin{align}
{\mathbb{I}}^{p_T^2}_S(\mu=p_T, \nu=p_T)=1\,, \qquad S^{(2)}(\mu=p_T,\nu=p_T)=0\,.
\end{align}
Note that the specific choice of $\mu=p_T$ is important for achieving this simple form of the RG since it eliminates the need to consider the diagonal terms in the mixing matrix. More generally, these would be required, but for LL resummation as considered in this paper, the particular path considered here suffices, as explained in more detail in \Sec{sec:N4_LL}.

This RGE is trivial to solve, and generates a single logarithm from the mixing
\begin{align}
S^{(2)}(\mu=p_T, \nu=Q)=
 \gamma_{\delta{\mathbb{I}} }  \log(p_T/Q)   {\mathbb{I}}^{p_T^2}_S(\mu=p_T, \nu=p_T) \,.
\end{align}
No additional logarithms are generated.  Since this is the case that applies to the soft function $S_{p_T^2}^{(2)}=p_T^2 S^{(0)}$, one can easily check using the known form of the two-loop soft function that there is indeed only a single logarithm at any $\nu$ scale for $\mu=p_T$. 

In an actual factorization formula, this single rapidity logarithm is then dressed by a Sudakov coming from running the hard function to the scale $\mu=p_T$, giving rise to a result of the form
\begin{align}
\gamma_{\delta{\mathbb{I}} }  \log(p_T/Q)   \exp \left[ -2 a_s \log^2\left( \frac{ p_T^2}{Q^2} \right)  \right]  H(Q) {\mathbb{I}}^{p_T^2}_S(\mu=p_T, \nu=p_T)\,.
\end{align}
 This provides quite an interesting structure, namely a single logarithm arising from $\nu$ evolution, which is then dressed by double logarithms from $\mu$ evolution. The $\nu$ and $\mu$ RGEs therefore completely factorize at leading logarithmic order. An identical structure was observed for the case of thrust in \cite{Moult:2018jjd}, however, there both the single logarithm and the tower of double logarithms arise from the $\mu$ RGE. 

\subsection*{Identity Operator ${\mathbb{I}}^\nu_S \left(\mu, \nu \right)$:}

Mixing with the rapidity identity operator ${\mathbb{I}}^\nu_S \left(\mu, \nu \right)$ gives rise to a more non-trivial rapidity flow at the scale $\mu=p_T$.
In this case, we get the RGE
\begin{align}
\nu \frac{d}{d\nu}\left(\begin{array}{c} S^{(2)} \\  {\mathbb{I}}^\nu_S \end{array} \right) &=  \left( \begin{array}{cc}0&\gamma_{\delta{\mathbb{I}} }\,  \\  0 & \gamma^\mu_S   \end{array} \right) \left(\begin{array}{c} S^{(2)} \\  {\mathbb{I}}^\nu_S \end{array} \right) \,, 
\end{align}
with the boundary conditions
\begin{align}
{\mathbb{I}}^\nu_S(\mu=p_T, \nu=p_T)=1\,, \qquad S^{(2)}(\mu=p_T,\nu=p_T)=0\,.
\end{align}
This RGE is is a specific case of the general RGE solved in \cite{Moult:2018jjd}. However, here we can solve it in two steps by first solving for the identity operator, and then plugging it in to the solution for the soft function. This will provide some insight into the structure of the final solution.

The solution of
\begin{align}
\nu \frac{d}{d\nu}  {\mathbb{I}}^\nu_S =\gamma^\mu_S  {\mathbb{I}}^\nu_S \equiv \tilde  \gamma \log\left(  \frac{\nu^2}{p_T^2} \right)  {\mathbb{I}}^\nu_S\,,
\end{align}
is easily found to be
\begin{align}
 {\mathbb{I}}^\nu_S=\exp\left( \frac{\tilde \gamma}{4} \log^2\left(  \frac{\nu^2}{p_T^2} \right) \right){\mathbb{I}}^\nu_S(\mu=p_T, \nu=p_T)\,.
\end{align}
The original soft function then satisfies the inhomogeneous equation
\begin{align}
\nu \frac{d}{d\nu} S^{(2)}=\gamma_{\delta{\mathbb{I}} }  \exp\left( \frac{\tilde \gamma}{4} \log^2\left(  \frac{\nu^2}{p_T^2} \right) \right){\mathbb{I}}_S^\nu(\mu=p_T, \nu=p_T)\,.
\end{align}
We can integrate this equation up to the scale $\nu=Q$ to find
\begin{align}
S^{(2)}(\mu=p_T, \nu=Q)=
 - \frac{\sqrt{\pi} \gamma_{\delta{\mathbb{I}} }}{ \sqrt{\tilde \gamma}}  \erfi\left[ \sqrt{\tilde \gamma}  \log(p_T/Q)  \right] {\mathbb{I}}^\nu_S(\mu=p_T, \nu=p_T) \,, 
\end{align}
where $\erfi$ is the imaginary error function, defined as 
\begin{align}
\erfi(z)=-i \erf(iz)\,.
\end{align}
The fact that the solution is not simply a Sudakov is quite interesting, and shows that the subleading power logarithms have a more interesting structure than at leading power. We expect that this structure will be quite common in the study of subleading power corrections to observables with rapidity evolution.
The $\erfi$ function satisfies the identify
\begin{align}
\frac{d}{dz}\erfi(z) = \frac{2}{\sqrt{\pi}} e^{z^2}\,.
\end{align}
It is perhaps quite intuitive that an integral of a Sudakov appears, since the rapidity identity operator is the integral of the leading power soft function. However, this is a new structure that has not previously appeared in subleading power calculations. It is amusing to note that a similar structure also appears in the calculation of Sudakov safe observables \cite{Larkoski:2015lea} due to the integration over a resummed result.

\section{The Energy-Energy Correlator in $\cN=4$ SYM}\label{sec:N4}

In this section we use our subleading power rapidity renormalization group to derive the leading logarithmic series at NLP for a physical observable, namely the EEC in $\cN=4$ SYM. We then compare our predictions with the recent calculation of \cite{Henn:2019gkr} to $\cO(\alpha_s^3)$ finding perfect agreement.

\subsection{Leading Logarithmic Resummation at Subleading Power}\label{sec:N4_LL}

In performing the resummation of subleading power logarithms for the EEC, we must clearly state several assumptions that are made, which we hope can be better understood in future work. Nevertheless, we believe that the fact that our result agrees with the calculation of  \cite{Henn:2019gkr} provides strong support for these assumptions. The goal of this paper has been to understand how far one can get in understanding the subleading power rapidity renormalization group using only symmetries and consistency arguments.  Using this approach, we found that at leading logarithmic order, there are two distinct identity operators with different renormalization group properties. To derive the structure of the resummed result for the EEC in the back-to-back limit, our approach will therefore be to match a linear combination of these two solutions to the know expansion of the EEC. To fix both coefficients requires two inputs, which we take to be the $\alpha_s$ and $\alpha_s^2$ leading logarithms. This then completely fixes our result, which can then be used to predict the coefficient of the $\alpha_s^3$ leading logarithm, for which we will find agreement with the calculation of \cite{Henn:2019gkr}. This approach should simple be viewed as a shortcut to a complete operator based analysis, which enables us to explore the general structure of the rapidity evolution equations at NLP, and show that they predict non-trivial behavior of the NLP series for a physical observable.  A more complete operator based analysis will be presented in a future paper.

Secondly, we must also assume that there exists a consistent factorization at subleading powers that does not have endpoint divergences. The presence of endpoint divergences in the factorization formula would violate our derivations based on the consistency of the RG. At this stage, both for the standard $\mu$ renormalization group at subleading power, and for the subleading power rapidity renormalization group considered here, this is still ultimately an assumption. In general, endpoint divergences are known to appear generically at next-to-leading logarithm, but have also been shown to appear even at LL in certain cases when fields with different color representations are involved (e.g. both quarks and gluons) \cite{Moult:2019uhz}. However, since in $\cN=4$ all fields are in the same representation, we work under the assumption that there are no endpoint divergences at leading logarithmic accuracy. We will see that this assumption is strongly supported by the fact that we are able to exactly reproduce to $\cO(\alpha_s^3)$ the highly non-trivial series that arises from the exact calculation of \cite{Henn:2019gkr}. However, we acknowledge that before our techniques can be more widely applied, it will be important to understand when endpoint divergences do occur in the rapidity renormalization group, and how they can be resolved. 

Therefore, working under the assumption of convergent convolutions for the subleading power factorization, all anomalous dimensions are fixed by symmetries, as described above, and the resummation of the subleading power logarithms is now a simple application of the renormalization group evolution equations derived in \Sec{sec:RRG_NLP}. To perform the resummation, one must resum logarithms in both $\mu$ and $\nu$. We choose to perform this resummation using the following evolution path
\begin{itemize}
\item Run the hard functions from $\mu=Q$ to $\mu=p_T$.
\item Run the soft functions in rapidity from $\nu=p_T$ to $\nu=Q$.
\end{itemize}
This path is the most convenient, since it avoids the need to perform any resummation of the rapidity anomalous dimensions \cite{Chiu:2012ir,Chiu:2011qc}. To run the hard function from $\mu=Q$ down to the soft scale, we use the evolution equations
\begin{align}
\mu \frac{d}{d\mu} H = \gamma_H H\,, \qquad \gamma_H =-8 a_s \log \left( \frac{ \mu^2}{Q^2} \right)\,,
\end{align}
Here and throughout this section, we will use $a_s=\alpha_s/(4\pi)C_A$ to simplify the notation.
The renormalization group equation for the hard function has the simple solution
\begin{align}
H(p_T)=H(Q) \exp \left[ -2 a_s \log^2\left( \frac{ p_T^2}{Q^2} \right)  \right]\,.
\end{align}
For the soft function evolution, we use the results derived in \Sec{sec:RRG_NLP} for the evolution of the two different types of rapidity identity operators. Since these rapidity identity functions cannot themselves mix due to different boost properties, the result at LL order is necessarily a linear combination of the two. For the first type of mixing, we have
\begin{align}
S_1^{(2)}(\mu=p_T, \nu=Q)=
 \gamma_{\delta{\mathbb{I}},1 }  \log(p_T/Q)   {\mathbb{I}}^{p_T^2}_S(\mu=p_T, \nu=p_T) \,,
\end{align}
where $ \gamma_{\delta{\mathbb{I}},1 }$ is the anomalous dimension for mixing into the  $ {\mathbb{I}}^{p_T^2}_S$ soft function, and for the second type, we have
\begin{align}
S_2^{(2)}(\mu=p_T, \nu=Q)=
 - \frac{\sqrt{\pi} \gamma_{\delta{\mathbb{I}},2 }}{ \sqrt{\tilde \gamma}}  \erfi\left[ \sqrt{\tilde \gamma}  \log(p_T/Q)  \right]  {\mathbb{I}}_S(\mu=p_T, \nu=p_T)\,,
\end{align}
with $\tilde \gamma =8 a_s$, and where $ \gamma_{\delta{\mathbb{I}},2 }$ is the anomalous dimension for mixing into the  $ {\mathbb{I}}^{\nu}_S$ soft function.
Our general prediction is then a linear combination of these two 
\begin{align}
\text{EEC}^{(2)}=\gamma_{\delta{\mathbb{I}},1 }  \log(p_T/Q)   \exp \left[ -2 a_s \log^2\left( \frac{ p_T^2}{Q^2} \right)  \right]   - \frac{\sqrt{\pi} \gamma_{\delta{\mathbb{I}},2 }}{ \sqrt{\tilde \gamma}}  \erfi\left[ \sqrt{\tilde \gamma}  \log(p_T/Q)  \right]\,.
\end{align}
Matching to the $a_s$ and $a_s^2$ coefficients from expanding the result of \cite{Belitsky:2013ofa,Belitsky:2013xxa,Belitsky:2013bja}(these coefficients are given in \Sec{sec:N4_expand}), we find that $\gamma_{\delta{\mathbb{I}},1 }=0$. We therefore find the simple result for the NLP leading logarithmic series to all orders in $a_s$ in $\cN=4$ SYM theory
\begin{align}
\text{EEC}^{(2)}=-\frac{\sqrt{\pi}a_s}{ \sqrt{2a_s}}  \erfi\left[ \sqrt{2 a_s} \log(1-z)  \right]  \exp\left[ -2 a_s \log(1-z)^2  \right]\,.
\end{align}
Interestingly, for the particular case of $\cN=4$, we find that the result only involves the operator ${\mathbb{I}}^\nu_S \left(\mu, \nu \right)$.
This result takes an interesting form, going beyond the simple Sudakov exponential \cite{Sudakov:1954sw} for the leading logarithms at leading power. We believe that this structure will appear somewhat generically in subleading power rapidity resummation.  It is interesting to note that up to the prefactor this particular structure is in fact a well known special function, called Dawson's integral, which is defined as
\begin{align}
D(x)=\frac{1}{2}\sqrt{\pi}e^{-x^2}\erfi(x)\,.
\end{align}
We can therefore write our answer for the NLP leading logarithmic series in the simple form
\begin{align}
\text{EEC}^{(2)}=-\sqrt{2a_s}~D\left[ \sqrt{2 a_s} \log(1-z) \right]\,.
\end{align}
Since the anomalous dimension is fixed by renormalization group consistency to be the cusp anomalous dimension (see \Eq{eq:rap_identity_summary}), we can rewrite this as
\begin{align}
\boxed{\text{EEC}^{(2)}=-\sqrt{2a_s}~D\left[ \sqrt{\frac{\gcusp}{2}} \log(1-z) \right]\,.}
\end{align}
This expression is a primary result of this paper. We will refer to this functional form as ``Dawson's Sudakov". We find it pleasing that despite the somewhat non-trivial functional structure, the double logarithmic asymptotics of the EEC in the back-to-back limit are still driven by the cusp anomalous dimension \cite{Korchemsky:1987wg} much like at leading power (see \Eq{eq:resformula_N4}), and as is expected physically. It would be extremely interesting to understand if the subleading power logarithms at next-to-leading power are driven by the collinear anomalous dimension, as is the case at leading power. We note that from \Eq{eq:N4_NLP_fixedorder} there is no constant term at NLP (it can easily be checked that this is true at any power), and therefore it is plausible that there is a simple generalization of this formula that also incorporates the next-to-leading logarithms.

\begin{figure}
\begin{center}
\includegraphics[width=0.75\columnwidth]{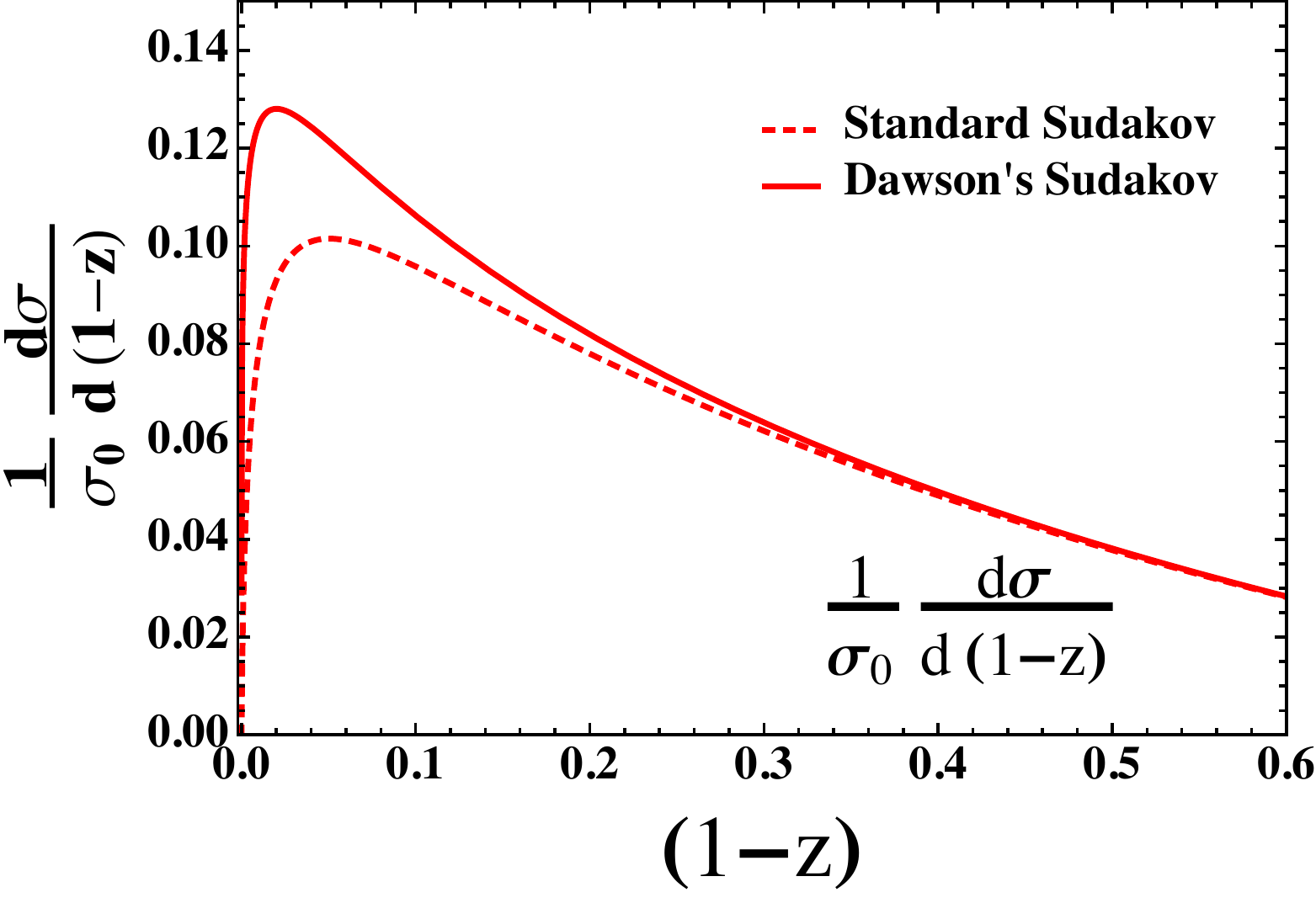}
\end{center}
\caption{A comparison of the standard Sudakov exponential which describes the all orders exponentiation of the leading power logarithms for the EEC with the subleading power Sudakov involving the imaginary error function ($\erfi$) derived in this section, which we refer to as Dawson's Sudakov.  Dawson's Sudakov exhibits a more peaked structure as $z\to 1$. A value of $\alpha_s=0.12$ was chosen to plot the numerical results.}
\label{fig:plot}
\end{figure}

In \Fig{fig:plot}, we compare the standard Sudakov with Dawson's Sudakov. Note that while $\erfi\left[ \sqrt{2 a_s} \log(1-z)  \right]$ diverges as $z\to 1$, this is overcome by the suppression from the Sudakov exponential. However, the behavior of the $\erfi$ leads to a more enhanced behavior of the distribution as $z\to 1$, as compared to a standard Sudakov.

It is also interesting to consider the Taylor expansion of our result in $a_s$. We find that it can be written as a remarkably simple series in terms of the double factorial
\begin{align}
\text{EEC}^{(2)}
&=\sum\limits_{n=0}^{\infty} \frac{(-1)^{n+1}}{(2n+1)!!} a_s^{n+1} \log((1-z)^2)^{2n+1}\,,
\end{align}
(here we have chosen to move factors of $2$ into the definition of the logarithm, but they can equally well be moved into the definition of $a_s$)
where we recall that the double factorial is defined as
\begin{align}
n!!=n(n-2)(n-4)\cdots .
\end{align}
Explicitly, the first few terms in the expansion are
\begin{align}\label{eq:EEC_RG_expand}
\text{EEC}^{(2)}&=-2 a_s \log(1-z) +\frac{8}{3}a_s^2 \log^3(1-z)-\frac{32}{15}a_s^3 \log^5(1-z) +\frac{128}{105}a_s^4 \log^7(1-z) \nn \\
&-\frac{512}{945}a_s^5 \log^9(1-z)+\frac{2048}{10395}a_s^6 \log^{11}(1-z)-\frac{8192}{135135}a_s^7 \log^{13}(1-z)+\cdots\,.
\end{align}
The presence of the double factorial as compared with the single factorial at leading power generates a more interesting series of rational coefficients.  In \cite{Dixon:2019uzg} the logarithms in the collinear limit of the EEC at each logarithmic order were written as simple infinite sums of factorials multiplied by polynomials in logarithms. It would be very interesting to understand if the subleading logarithms at NLP in the back-to-back limit can also be written as generalized double factorial sums.

It will be important to understand if this structure persists in QCD. The results in QCD are only known to $a_s^2$ \cite{Dixon:2018qgp,Luo:2019nig}, therefore, without performing a more complete operator analysis, these results can be used to fix our prediction as a linear combination of our two RG solutions, but do not enable a non-trivial check, which is particularly important due to the possible presence of endpoint divergences. While we will perform an operator based analysis in a future publication, we find that it interesting to conjecture a result. Based on our intuition from the case of the thrust observable \cite{Moult:2019uhz}, we expect that we are most likely to avoid endpoint divergences for the case of the EEC in Higgs decays to gluons in pure Yang-Mills. Under the assumption that there are no endpoint divergences, we can fix an ansatz in terms of our two RG solutions to be
\begin{align}
\left.\text{EEC}^{(2)}\right|_{\text{Yang-Mills}}=2 a_s \log(1-z)   \exp \left[ -2 a_s \log^2\left(1-z \right)  \right]   -4\sqrt{2a_s}~D\left[ \sqrt{\frac{\gcusp}{2}} \log(1-z) \right]\,,
\end{align}
which takes a slightly more complicated form than its $\cN=4$ counterpart, since it involves both types of rapidity identity operators. Unfortunately, unlike for the case of the EEC in $\cN=4$, there is no $a_s^3$ result to which we are able to compare this result. 
It will interesting to verify or disprove this conjecture using a complete operator analysis, which we leave to future work. This will provide insight into the presence (or lack of) endpoint divergences in this particular case.

It would also be extremely interesting to understand how to derive this result directly from the four point correlator, or from the light ray OPE \cite{Kravchuk:2018htv,Kologlu:2019bco,Kologlu:2019mfz}. From the point of view of the four point correlator, the back-to-back limit corresponds to the so called double light cone limit. This has been used to study the EEC in \cite{Korchemsky:2019nzm}, has been studied in gauge theories in \cite{Alday:2010zy,Alday:2013cwa}, and has been studied in more general conformal field theories in \cite{Alday:2015ota}. 

\subsection{Comparison with Direct Fixed Order Calculation to $\cO(\alpha_s^3)$}\label{sec:N4_expand}

As mentioned earlier, we have chosen to perform the resummation for the EEC in $\cN=4$ since we can directly compare with the remarkable calculation of the EEC for arbitrary angles using the triple discontinuity of the four point correlator \cite{Henn:2019gkr} (The result to $\cO(\alpha_s^2)$ in $\cN=4$ was calculated in \cite{Belitsky:2013ofa,Belitsky:2013xxa,Belitsky:2013bja}), and exploiting the large amount of information known about its structure, see e.g. \cite{Eden:2011we,Eden:2012tu,Drummond:2013nda,Korchemsky:2015ssa,Bourjaily:2016evz}. This calculation of the EEC provides extremely valuable data for understanding the structure of kinematic limits at subleading power, both for the particular case considered here, and beyond.

Although we will be interested in the expansion in the $z\to 1$ limit, it is interesting to understand what parts of the full angle result the back-to-back limit is sensitive to at subleading powers, so we briefly review the structure of the result of \cite{Henn:2019gkr}.
The result of \cite{Henn:2019gkr} is written as
 \begin{align}\label{defF}
 F(\z) \equiv 4 \z^2 (1-\z) \EEC (\z) \,,
 \end{align}
where at NNLO, 
 \begin{align}  \label{resultFatNNLO}
F_{\rm NNLO} (\zeta) &= f_{\rm HPL} (\zeta) +   \int_{0}^1 d \bar z  \int_{0}^{\bar z } dt \, \frac{\z-1}{t(\z-\bz)+(1-\z)\bz} \nt 
& \times \left[  R_1 ( z, \bar z )  P_1 (z, \bar z ) + 
  R_2 ( z, \bar z )  P_2 (z, \bar z ) \right] \,,
\end{align} 
with
 \begin{align} 
  R_1 = \frac{z \bar z }{ 1 - z - \bar z}\, , \quad  R_2 = \frac{z^2 \bar z }{ (1-z)^2  ( 1 - z \bar z)}\,.
 \end{align}
 Here $P_1$ and $P_2$ are weight three HPLS in $z$ and $\bar z$. These two fold integrals are believed to be elliptic, and so we will refer to them as elliptic contributions. The term $f_{\rm HPL} (\zeta)$ is expressed in terms of harmonic polylogarithms. The leading power asymptotics in the back-to-back limit are described entirely by $ f_{\rm HPL}$. At NLP, we require also the elliptic contributions, showing that subleading powers probe more of the structure of the result.  This in turn makes the agreement with our result derived from the RG more non-trivial.

The expansion of $f_{\rm HPL} (\zeta)$ can be performed straightforwardly, and produces only $\zeta$ values, $\log^n(2)$, and $\text{Li}_n(1/2)$. On the other hand, the expansion of the elliptic contribution is more non-trivial, and leads to a more complicated set of constants. To compute the result, we expanded under the integral sign, and integrated using HyperInt \cite{Panzer:2014caa}. This produced polylogarithms of sixth roots of unity up to weight 5. These were reduced using results from  \cite{Henn:2015sem} to a basis of constants. The final result involves several non-zeta valued constants which were guessed using hints for the classes of numbers that should appear in the answer from \cite{Broadhurst:1998rz,Fleischer:1999mp,Davydychev:2000na} and reconstructed using the PSLQ algorithm. We found that the elliptic piece could be expressed as 
\begin{align}
\frac{\text{Elliptic}_{\text{NLP}}}{2}&=2\frac{L^5}{5!}+\frac{1}{2}\zeta_2 \frac{L^3}{3!}  +\frac{3}{4}\zeta_3 \frac{L^2}{2!}+ \left(-\frac{67}{32}+\frac{3}{4}\zeta_2+\frac{9}{4}\zeta_3 -\frac{7}{4} \zeta_4\right) L\nn \\
&+\frac{85}{32}-\frac{49}{16}\zeta_2+\frac{87}{8}\zeta_3 +\frac{37}{4}\zeta_4 +\frac{3}{4}\zeta_2\zeta_3+\frac{5}{2}\zeta_5-\frac{611}{108}\zeta_4 \sqrt{3} \pi+6\sqrt{3}I_{2,3}\,.
\end{align}
Here $I_{2,3}$ is a higher weight Clausen function \cite{Broadhurst:1998rz} 
\begin{align}
I_{2,3}=\sum\limits_{m>n>0}\frac{\sin\left(\frac{\pi(m-n)}{3}\right)}{m^{b-a}n^{2a}}\,.
\end{align}
While the leading power result has uniform trascendental weight when expanded in the back-to-back region, this is no longer true at subleading power.

Collecting all the terms from both the elliptic and HPL contributions, we find (in our normalization) the following expression for the leading power suppressed logarithms up to $\cO(a_s)^3$
\begin{align}
 &\text{EEC}^{(2)}=-2a_s \log(1-z) \nn \\
 &+a_s^2 \left[\frac{8}{3} \log^3(1-z) +3 \log^2(1-z) +(4+16\zeta_2)\log(1-z)+(-12-2\zeta_2+36 \zeta_2\log(2)+5\zeta_3) \right]\nn \\
 &+a_s^3 \left[-\frac{32}{15} \log^5(1-z)-\frac{16}{3}\log^4(1-z)-\left(  \frac{8}{3}+24 \zeta_2 \right)\log^3(1-z) +(4-36\zeta_2-50\zeta_3)\log^2(1-z) \right. \nn \\
 &\left.  -\left(\frac{131}{2}+4\zeta_2+372 \zeta_4+12\zeta_3   \right) \log(1-z) + \text{const}  \right]\,,
\end{align}
where
\begin{align}
\text{const}&=-\frac{3061}{2}\zeta_5  -96\zeta_2 \zeta_3 -\frac{4888}{27}\zeta_4 \pi \sqrt{3} +192 \sqrt{3} I_{2,3}+1482 \zeta_4 \log(2) -256 \zeta_2 \log^3(2) \nn \\
&-\frac{64}{5}\log^5(2)
+1536 \text{Li}_5\left( \frac{1}{2} \right) -544 \zeta_4 +192 \zeta_2 \log^2(2)+16 \log^4(2)+384 \text{Li}_4\left( \frac{1}{2} \right) \nn \\
& -288 \zeta_2 \log(2)+158 \zeta_3 +55 \zeta_2 +\frac{533}{2}\,.
\end{align}
Extracting out the leading logarithmic series
\begin{align}
\left. \text{EEC}^{(2)}\right|_{\text{LL}}=-2a_s \log(1-z) +\frac{8}{3}a_s^2 \log^3(1-z) -\frac{32}{15}a_s^3 \log^5(1-z)\,,
\end{align}
we find that this result agrees exactly with the result derived from the subleading power renormalization group given in \Eq{eq:EEC_RG_expand}! Note that due to the matching of our RG predictions to the fixed order results, as was explained above, it is really only the $a_s^3$ coefficient that is a prediction. However, this agreement is highly non-trivial, since it probes both the elliptic and polylogarithmic sectors of the full result, and therefore we believe that it provides strong support that our subleading power renormalization group evolution equation is correct. The next-to-leading logarithms also have relatively simple rational coefficients, 
\begin{align}
\left. \text{EEC}^{(2)}\right|_{\text{NLL}}=3a_s^2 \log^2(1-z) -\frac{16}{3}a_s^3 \log^4(1-z)\,,
\end{align}
and provide data for understanding the structure of subleading power resummation beyond the leading logarithm. It would be interesting to derive them directly from the renormalization group approach.

It would be interesting to better understand the structure of the numbers appearing in the expansion of the EEC both in the collinear and back-to-back limits, and the functions appearing in the full angle result. This could ultimately allow the result to be bootstrapped from an understanding of these limits, in a similar manner to the hexagon bootstrap for $N=4$ SYM amplitudes \cite{Dixon:2011pw,Dixon:2014iba,Caron-Huot:2016owq,Dixon:2016nkn,Caron-Huot:2019vjl}. However, the presence of elliptic functions makes this seem like a daunting task, unless more information is known about their structure.

\section{Conclusions}\label{sec:conc}

In this paper we have shown how to resum subleading power rapidity logarithms using the rapidity renormalization group, and have taken a first step towards a systematic understanding of subleading power corrections to observables exhibiting hierarchies in rapidity scales. Much like for the virtuality renormalization group at subleading power, the rapidity renormalization group at subleading power involves a non-trivial mixing structure. Using the consistency of the RG equations combined with symmetry arguments, we were able to identify the operators that arise in the mixing,  which we termed ``rapidity identity operators", and we derived their anomalous dimensions. We believe that these operators will play an important role in any future studies of the rapidity renormalization group at subleading power, and are the key to understanding its structure. 

To illustrate our formalism, we resummed the subleading power logarithms appearing in the back-to-back limit of the EEC in $\cN=4$ SYM. This particular observable was chosen since the full analytic result is known to $\cO(\alpha_s^3)$ from the remarkable calculation of \cite{Henn:2019gkr}. We found perfect agreement between our result derived using the renormalization group, and the expansion in the back-to-back limit of the calculation of \cite{Henn:2019gkr}, which provides an extremely strong check on our results. The analytic form of the resummed subleading power logarithms takes an interesting, but extremely simple form, being expressed in terms of Dawson's integral with argument related to the cusp anomalous dimension. We called this structure ``Dawson's Sudakov". We expect this structure to be generic at subleading power for rapidity dependent observables, much like the Sudakov exponential is at leading power. 

Since this represents the first resummation of subleading power rapidity logarithms, there are many directions to extend our results, as well as to better understand the structures that we have introduced in this paper. First, although we have arrived at the structure of the leading logarithms using symmetry and consistency arguments, it would be interesting to use a complete basis of \scetii ~operators to derive the operator structure of all the subleading power jet and soft functions in \scetii, and perturbatively compute their anomalous dimensions. Second, to go beyond LL, it will be necessary to better understand the structure of the momentum convolutions appearing in the subleading power factorization. We expect that at subleading power this is best done in momentum space using the formalism of \cite{Ebert:2016gcn}. Finally, we also expect that away from $\cN=4$ SYM theory, divergent convolutions will appear even at LL order, as occurs in \sceti, and so it will be important to understand when these occur, and how they can be overcome.

On the more formal side, it will be interesting to understand how to extract the subleading power logarithms directly from the four point correlation function, following the approach of  \cite{Korchemsky:2019nzm}, or using the light ray OPE \cite{Kravchuk:2018htv,Kologlu:2019bco,Kologlu:2019mfz}. While there have been some studies of the double light cone limit in conformal field theories \cite{Alday:2010zy,Alday:2015ota}, further studies of this limit in conformal field theories could provide insight into behavior of phenomenological interest in QCD, and the EEC provides an example that is of both formal and phenomenological interest.

It will also be important to apply our formalism to observables of direct phenomenological interest, such as the EEC in QCD, the color singlet $p_T$ distribution in hadron colliders, or to the study of power suppressed logarithms appearing in the Regge limit, which can also be formulated in terms of the rapidity renormalization group \cite{Rothstein:2016bsq,Moult:2017xpp}.  Our work represents the first step in extending recent successes in understanding the structure of subleading power infrared logarithms to subleading power rapidity logarithms, and we hope that this will allow for a much wider set of phenomenologically important applications.

\begin{acknowledgments}
We thank Martin Beneke, Sebastian Jaskiewicz, Robert Szafron, Iain Stewart, Frank Tackmann, Johannes Michel, Markus Ebert, David Simmons-Duffin, Cyuan-Han Chang, Lance Dixon, Johannes Henn, and HuaXing Zhu for useful discussions. We thank Vladimir Smirnov for providing us with reduction tables for polylogarithms of roots of unity. This work was supported in part by the Office of Nuclear Physics of the U.S.
Department of Energy under Contract No. DE-SC0011090, and by the Office of High Energy Physics of the U.S. Department of Energy under Contract No. DE-AC02-76SF00515. This research received
funding from the European Research Council (ERC) under the European
Union's Horizon 2020 research and innovation programme
(grant agreement No. 725110), ``Novel structures in scattering amplitudes".
\end{acknowledgments}

\bibliography{subRGE}{}
\bibliographystyle{jhep}

\end{document}